\documentclass[preprints,article,accept,pdftex,oneauthor]{Definitions/mdpi} 

\firstpage{1} 
\pubvolume{1}
\issuenum{1}
\articlenumber{0}
\pubyear{2023}
\copyrightyear{2023}
\datereceived{ } 
\daterevised{ } 
\dateaccepted{ } 
\datepublished{ } 
\hreflink{https://doi.org/} 

\DeclareMathOperator{\sign}{sign}
\newcommand{\Mathematica}{\textit{Mathematica\textsuperscript{\resizebox{!}{0.8ex}{\textregistered}}}}
\def\8{\infty}
\def\oh{\frac{1}{2}}
\def\ot{\frac{1}{3}}
\def\oq{\frac{1}{4}}
\def\tt{\frac{2}{3}}
\def\ft{\frac{4}{3}}
\def\tq{\frac{3}{4}}

\newcommand*{\I}{\imath}%

\def\eps{\epsilon}

\def\dal{\partial_{\alpha}}
\def\dbe{\partial_{\beta}}

\def\const{\textit{ const }}
\def\undertext#1{\vtop{\hbox{#1}\kern 1pt \hrule}}
\def\ra{\rightarrow}
\def\Ra{\Rightarrow}

\def\MO#1{O\left(#1\right)}
\def\abs#1{\left| #1\right|}
\def\tens#1{\lVert #1\rVert}
\def\pd#1{\partial_{#1}}
\def\VEV#1{\left\langle #1\right\rangle}

\def\tr{\hbox{tr}\,}
\def\diag#1{\hbox{diag}\left(#1\right)}

\def\dbyd#1#2{\frac{d#1}{d#2}}

\def\pbyp#1#2{\frac{\partial#1}{\partial#2}}

\def\bea{\begin{eqnarray} & &}
\def\eea{\end{eqnarray}}

\let\oldexp\exp
\renewcommand{\exp}[1]{\oldexp\left(#1\right)}

\def\IINT#1{\int_{#1}\int_{#1}}

\def\et{{\mathcal E}}
\def\CL{Clebsch}

\def\KO{Kolmogorov}
\def\RE{\textbf{Rey}}
\def\NS{Navier-Stokes}

\def\BS{Biot-Savart}

\def\CVS{\textit{CVS}}

\def\val{v_{\alpha}}
\def\vbe{v_{\beta}}
\def\vga{v_{\gamma}}

\def\ral{r_{\alpha}}
\def\rbe{r_{\beta}}

\def\XXint#1#2#3{{\setbox0=\hbox{$#1{#2#3}{\int}$}
     \vcenter{\hbox{$#2#3$}}\kern-.5\wd0}}

\newcommand{\bZ}{\mathbb{Z}}
\newcommand{\bS}{\mathbb{S}}
\newcommand{\bR}{\mathbb{R}}
\newcommand{\bD}{\mathbb{D}}
\newcommand{\bT}{\mathbb{T}}

\DeclareMathOperator*{\argmin}{arg\,min}

\newcommand{\pct}[2]{
\begin{figure}
    \includegraphics[width=\textwidth]{#1}
    \caption{#2}
    \label{fig::#1}
\end{figure}
}

\Title{TOPOLOGICAL VORTEXES, ASYMPTOTIC FREEDOM AND MULTIFRACTALS}

\TitleCitation{Title}

\Author{Alexander Migdal $^{1,\dagger,\ddagger}$\orcidA{}}

\AuthorNames{Alexander Migdal}

\AuthorCitation{Migdal, A.~A.}

\address{
$^{1}$ \quad Department of Physics, New York University Abu Dhabi,
            Saadiyat Island, 
            Abu Dhabi,
            PO Box 129188, 
           Abu Dhabi,
            United Arab Emirates; am10485@nyu.edu}

\corres{Correspondence: sasha.migdal@gmail.com}

\abstract{We study the Kelvinons: monopole ring solutions to the Euler equations, regularized as the Burgers vortex in the viscous core.
 There is finite anomalous dissipation in the inviscid limit. However,  in the anomalous Hamiltonian, some terms are growing as logarithms of Reynolds number; these terms come from the core of the Burgers vortex. In our theory, the turbulent multifractal phenomenon is similar to asymptotic freedom in QCD, with these logarithmic terms summed up by an RG equation. The small effective coupling does not imply small velocity; on the contrary, velocity is large compared to its fluctuations, which opens the way for a quantitative theory. 
 In the leading order in the perturbation theory in this effective coupling constant, we compute running multifractal dimensions for high moments of velocity circulation, in good agreement with the data for quantum Turbulence and available data for classical Turbulence. The logarithmic dependence of fractal dimensions on the loop size comes from the running coupling in anomalous dimensions. This slow logarithmic drift of fractal dimensions would be barely observable at Reynolds numbers achievable at modern DNS.}


\begin{document}
\setlength{\headheight}{20.0pt}
\setcounter{section}{-1} 

\section{Introduction}

In 1948 Burgers discovered his vortex solution to the \NS{} equations \cite{BURGERS1948}. This exact stationary solution with cylindrical geometry had an anomalous dissipation (a finite dissipation at vanishing viscosity).

It should have been a breakthrough in the theory of Turbulence. Instead, a phenomenological K41 theory by Kolmogorov and Obukhov \cite{Kolm41} dominated the turbulence studies for the next 80 years, while the Burgers' discovery was almost forgotten.  

While qualitatively describing some important turbulence features, the K41 scaling laws led the microscopic theory to a dead end.

The main question: How is the inviscid \NS{} theory different from the Euler theory was unanswered, nor were the other nascent questions like what is the microscopic mechanism of the spontaneous stochasticity?

In the last few years, there has been some progress in understanding the role of the singular topological solutions of the Euler equations, with Burgers solutions resolving singularities. This progress was inspired by the geometric approach to Turbulence initiated in the 80-ties and 90-ties \cite{M88, AM89, M93, TSVS}, leading to the area law prediction \cite{M93}.

This prediction was verified recently in DNS \cite{S19, S21}, which triggered a flux of new studies in the geometric theory of Turbulence.

This work was recently revised and summarized in a review paper \cite{M23PR}.

Based on initial comments, the details of this theory will take time for the turbulent community to process. The theory will likely need to be further explained, discussed, and split into smaller sets that can be advanced separately.

There are many subtleties to be clarified, some minor corrections to be made, and some questions to be answered. Also, there are more data to compare with the Kelvinon theory and the Loop equation solutions.

\section{Weak and strong phases of Turbulence}

Before we go into details of the theory of singular topological Euler solutions and velocity fluctuations in their vicinity, it is worth reminding the readers that we are talking about a \textbf{different phase of Turbulence}, not the one corresponding to Wylde's diagrams and attempts of summing up these diagrams by some "renormalization group."

These diagrams describe the terms of expansion of the correlation functions of a very viscous velocity field in inverse powers of viscosity. However, we are interested in the opposite limit of viscosity going to zero. This limit is known to be singular: the vortex sheets and lines arising in this limit cannot be expanded in the inverse powers of viscosity. 

We study these singular solutions of the Euler equations and see how the infinitesimal viscosity resolves these singularities rather than trying to extrapolate the viscous flow to a turbulent phase. So to say, we meet the inviscid limit of the \NS{} equation face to face instead of walking back far away and looking into a binocular. 

Yakhot and Zakharov described the weak Turbulence \cite{YZ93} by assuming that \CL{} field varied in a 2D plane $\bR^2$. In that case, there were particles propagating as waves in 3D space and obeying the conservation of the number of particles. Under some further model approximations, these authors derived the K41 spectrum from the kinetic equation in a one-loop term of the perturbation theory.

This model made sense then, as the K41 laws were observed in weak Turbulence in real and numerical experiments.

There is, however, a strong evidence in modern DNS\cite{S19, SY21, SY22} that there are one or more phase transitions as the Reynolds number increases.

In our approach \cite{M23PR}, there is a \textbf{compactification phase transition}  $\bR^2\Ra \bS^2$ in the target space of the \CL{} variables.
In this new phase, the \CL{} variables are rigid rotators (or classical spin variables) in the Hamiltonian mechanics. They are \textbf{confined} by unbroken gauge invariance under area-preserving diffeomorphisms in two-dimensional target space.

Note that it is NOT the volume-preserving transformations of incompressible fluid dynamics; those transformations in physical space $\bR^3$ do not correspond to any gauge invariance of the \NS{} equations, unlike these area-preserving diffeomorphisms in the target space $\bS^2$.

Our gauge transformations keep velocity and vorticity invariant while changing the \CL{} variables. Geometrically, these variables are points on a sphere $\bS^2$, and area-preserving diffeomorphisms are local transformations of variables with unit Jacobian.

For the Quantum and Classical Gravity experts, our compact \CL{} variables constrained to a sphere $\bS^2$ are similar to the Ashtekar variables in Hamiltonian Gravity theory \cite{AshtekarVar}. Our singular topological solution also has analogs in the Gravity theory: these are so-called KLS domain walls bounded by strings \cite{Vol15, ZH20}. Our Loop Equation \cite{M93, M23PR} was inspired by the QCD loop equation \cite{M83} and bears some resemblance to the loop quantum gravity \cite{LoopGravity}.

The common thread of all these approaches is to treat nonlinear dynamics as dynamical geometry, aiming to discover nontrivial strong coupling phases with some topological invariants. The turbulence problem, the oldest of these nonlinear problems, does not need these analogies. The topology and dynamics we have found are shockingly simple compared to QCD and Quantum Gravity.

We have found that this confining phase is very different from the weak Turbulence, where the velocity fluctuations grew with a viscosity decrease.
A singular velocity field describes a monopole ring in a confined or strong phase. The \CL{} field is like a  potential of the magnetic monopoles; the vorticity is a single-valued function, but the velocity field is multivalued.

The velocity field has a gap at a disk-like surface bounded by this monopole ring; at the ring, velocity diverges, and the Euler equation does not apply at the surface, including the ring's core. The \NS{} equation describes the velocity in these singular regions.

The dynamical variables outside the singularities are components of the \CL{} field $S \in \bS^2$. The logarithmic divergences arise in the \textbf{classical} theory by minimizing the Euler Hamiltonian over \CL{} variables. The fluctuations around these classical solutions are small in the inviscid limit.

We arrive at the RG equations and then at asymptotic freedom (these fluctuations are inversely proportional to the logarithm of the Reynolds number).

Neither the physical mechanism of our asymptotic freedom nor the mathematical structure resembles the old RG equations in the weak turbulence phase.
The definition of effective coupling is different; the dynamical equations are different (classical in our case, fluctuating in the old RG case).

The most important difference is that our beta function is \textbf{exact}, not some model or perturbative approximation.

With these comments in mind, let us proceed to our theory.

\section{Trying to bend the Burgers cylindrical vortex into a torus}

Let us start with the Burgers vortex solution of the \NS{} equation:
\begin{subequations}
\begin{eqnarray}\label{BurgersVortex}
    && \vec v =\left\{-a x- g(r) y, -b y + g(r) x, c z + d\right\};\\
    && \vec \omega = \left\{0,0,\frac{c  \Gamma_B  e^{-\frac{c r^2}{4 \nu }}}{8 \pi  \nu }\right\};\\
    && g(r) = \frac{\left(1-e^{-\frac{c  r^2}{4 \nu }}\right)\Gamma_B  }{2 \pi  r^2};\\
    && r = \sqrt{x^2 + y^2};\\
    && a = b = c/2 ;
\end{eqnarray}
\end{subequations}

This solution describes an infinite cylinder with vorticity decaying as a Gaussian with the width
\begin{eqnarray}
    w = \sqrt{2 \nu/c};
\end{eqnarray}
This constant $c$ represents the strain $\hat S$ in the direction of the symmetry axis.
\begin{eqnarray}
   && c = \vec t \cdot \hat S \cdot \vec t;\\
   && \hat t = \{0,0,1\}; \\
   \label{strain}
   && \hat S = \tens{ S_{\alpha\beta} } =  \frac{\pd\beta \val + \pd\alpha \vbe}{2} = \diag{-a,-b,c}
\end{eqnarray}

The parameter $\Gamma_B$  represents the circulation in the closed loop surrounding this axis far from its center. By the Stokes theorem, this circulation reduces to an area integral over the surface bounded by this loop, which tends to $\Gamma_B$ in the limit when the radius is much larger than the width $w$ of the vorticity core.

This relation can be directly verified by integrating Burgers velocity around the contour and neglecting exponential terms in $g(r)$.

It is also straightforward to compute anomalous dissipation
\begin{eqnarray}
    \et  =  \nu \int d^3r \vec \omega^2 =  \frac{L c \Gamma_B^2}{8 \pi}
\end{eqnarray}
where $L \ra \infty$ is the length of the cylinder.

Let us try to compactify the Burgers vortex by bending a cylinder into the torus.
Consider a circular vortex line with a radius $R$ much larger than the thickness $w$ of the viscous core.

Locally, at the distances from the core $r \sim w$, the torus is equivalent to the cylinder, up to higher order terms in $w/R$. 

At larger distances, the vorticity is equivalent to the delta function in the cross-section plane, and velocity is purely potential.
It adds up from  a linear term and a singular term
\begin{eqnarray}
&&\vec \omega(\vec r) \ra \Gamma_B \vec t \delta(x) \delta(y);\\
   &&\vec v(\vec r) \ra \left(-a x,-b y, c z\right) + \frac{\Gamma_B}{2 \pi} \frac{ \left(-y,x,0\right)}{x^2 + y^2};
\end{eqnarray}

An important property of the Burgers solution is that the strain tensor \eqref{strain} is constant and has no singularities in the inviscid limit, unlike the rotational part of velocity.
This constant strain represents the local value of the nonsingular Euler strain outside the vortex core, continued inside the tube..

The invariant formula for the inviscid limit of vorticity would be the line integral (with $l$ being the length of the line)
\begin{eqnarray}
    &&\vec \omega(\vec r) = \oint_C d \vec C(l) \Gamma_B(l) \delta^3(\vec r - \vec C(l));\\
    && |\vec C'(l)| =1
\end{eqnarray}
When the point $\vec r$ approaches some point $\vec C(l_0)$ at the loop, the loop integration cancels the delta function, and one recovers the Burgers delta function in the $x y$ plane.
\begin{eqnarray}
    &&\vec \omega(\vec r) \ra \vec C'(l_0) \Gamma_B(l_0) \delta(x) \delta(y);\\
    &&\vec C'(l_0) = (0,0,1);
\end{eqnarray}

In the linear vicinity of a point $\vec C(l_0)$ at this loop, we use the Burgers solution with $z = l-l_0$ to find the derivatives
\begin{subequations}\label{velocityOnLoop}
\begin{eqnarray}
    &&\partial_l \Gamma_B =0;\\
    &&\partial_l\vec v\left(\vec C(l)\right) = \hat S\left(\vec C(l)\right) \cdot \vec C'(l) ;
\end{eqnarray}
\end{subequations}
The velocity field here is taken at the loop $C$, which is a center of the Burgers core.

There is an important boundary condition for the strain tensor at the loop
\begin{eqnarray}\label{EigenvalueCondition}
    && \hat S \left(\vec C(l)\right) \cdot \vec C'(l) = c(l) \vec C'(l);
\end{eqnarray}
In general, the eigenvalue $c(l)$ depends on the point $l$ at the loop, as the strain could be a function of coordinates and the loop changes direction.
This eigenvalue must be positive.

This boundary condition is an analog of the \CVS{} conditions \cite{M23PR} (confined vortex surface) relating the vortex sheet shape and the boundary value of the strain.
In the case of the vortex sheet, this relation imposed restrictions on a surface shape, as the Euler strain did not have any free parameters to adjust. 

In the case of the loop, these restrictions can be treated as extra boundary conditions on the more complex Euler flow outside the tube, as we shall see in the next Sections.

These restrictions are quite strong. In the case of a curved loop we are looking for, the constant strain cannot have the loop tangent vector $\vec C' (l)$ as its eigenvector. 

Unlike the Burgers formulas, these relations \eqref{velocityOnLoop} are parametric invariant and do not depend on the coordinate frame.
This invariance makes them correct generalizations of Burger's solution to an arbitrary smooth loop.

These equations can be readily integrated
\begin{eqnarray}
    &&\Gamma_B(l) = const;\\
    \label{VelocityStrainIntegral}
    && \vec v\left(\vec r\in C\right)= \int^{\vec r} \hat S \left(\vec r'\right) \cdot d \vec r'
\end{eqnarray}

We found a parametric invariant vector integral of the strain along the loop. Another way to derive this formula is as follows:
\begin{eqnarray}
    &&\vec v(\vec r \in C) = \int_C^{\vec r} d \vec r'\cdot \vec \nabla \otimes  \vec v\left(\vec r'\right) =\nonumber\\
    && \int_C^{\vec r} d \vec r'\cdot \hat S(\vec r') + \int_C^{\vec r} d \vec r' \times \vec \omega(\vec r')
\end{eqnarray}
The second term here vanishes because the vorticity at the loop $\vec \omega(\vec r')$ is aligned with its tangent vector $d \vec r'$.

From this relation, by integrating by parts, one can derive the following exact relation for the circulation over the large loop $C$
\begin{eqnarray}
    \Gamma_C = \oint_C d \ral \val = - \oint_C \ral d \val = - \oint_C \ral   S_{\alpha\beta} d\rbe 
\end{eqnarray}
In virtue of the eigenvalue equation, this is also equivalent to
\begin{eqnarray}
    \Gamma_C = -\oint_C d l  \vec C'(l) \cdot \vec C(l) c(l)
\end{eqnarray}

The generalization of Burger's anomalous dissipation is also straightforward \cite{M23PR}:
\begin{eqnarray}\label{AnomalousDissipation}
    \et = \frac{\Gamma_B^2}{8\pi} \oint_C d l \vec C'(l)\cdot \hat S \cdot \vec C'(l) = \frac{\Gamma_B^2}{8\pi} \oint_C d l c(l)
\end{eqnarray}

The problem we are facing is related to the strain. As we have seen, it cannot be a constant tensor (its highest eigenvector must be equal to the local direction of the loop everywhere).

The generic Euler (singular) velocity field, corresponding to the above singular vorticity line, reads
\begin{eqnarray}\label{SingularLineVelocity}
    \vec v(\vec r) \stackrel{?}{=} \vec \nabla \Phi - \frac{\Gamma_B}{4 \pi} \vec \nabla \times \oint_C\frac{d \vec r'}{\abs{\vec r - \vec r'}}
\end{eqnarray}

The harmonic potential $\Phi(\vec r)$ must be such that the strain at the loop
\begin{eqnarray}
    S_{\alpha\beta}\left(\vec C(l)\right) \stackrel{?}{=} \dal \dbe \Phi\left(\vec C(l)\right) 
\end{eqnarray}
has the local tangent vector $\vec C' (l)$ as its main eigenvector at every point on the loop. 

This main eigenvalue $c$ must be positive.
Two lower eigenvalues $-a,-b$ do not have to be equal, as a further study of the Burgers vortex revealed. As Moffat et al. \cite{MKO94} have found, the vortex solution for a general non-axisymmetric strain tends to the symmetric Burgers solution in the turbulent limit $|\Gamma_B| \gg \nu$. 

We are only interested in that limit; therefore, we can skip the requirement of equal lower eigenvalues.

So, is this it? A weak Euler flow regularized by a Burger vortex core?
Not so fast.

This solution would not be valid for an arbitrary smooth loop $C$ because, in general, the eigenvalue condition will not hold.

We are unaware of any theorems that would prove the existence of the harmonic potential with the prescribed main eigenvector of its Hessian $S_{\alpha\beta}$ on a closed loop $C$ in space.
Presumably, such a harmonic potential does not exist for an arbitrary (or even smooth) loop.

\section{Matching principle and anomalies in the Euler Hamiltonian}

The stationary solution we are looking for must minimize the Hamiltonian
\begin{eqnarray}
    H = \int_{\bR^3} \frac{\vec v^2}{2}
\end{eqnarray}

In our case of the singular velocity field, we have to split this energy integral into two parts: inside and outside a thin tube $\mathcal T$ surrounding the loop $C$. The radius $R$ of a local cross-section of the boundary $\partial \mathcal T$ of this tube must be much larger than Burger's thickness $w$ but much smaller than the local curvature radius $R_C$ of the loop $C$.

Under these conditions, the inside of the tube is described by a cylindrical Burgers vortex, while the outside is some Euler flow in the remaining space $\mathcal G = \bR^3 \setminus \mathcal T$. This remaining space has the topology of the full torus, the same as $\mathcal T$.
We understand the space $\bR^3$ is compactified as a sphere $\bS^3$ by including the infinity.

At the surface of the tube and its vicinity, for the whole range $ w \ll R \ll R_C$, there must be a match of the inside Burgers flow up to higher order corrections in $w/R$ with the (yet unspecified) Euler flow up to $R/R_C$ corrections.

This requirement is a particular case of the matching principle, which we suggested first for the vortex sheets, and then for the vortex lines (see \cite{M23PR} and references to earlier work within).

Thus, the Hamiltonian can be written as the sum of two terms
\begin{eqnarray}
    H = \int_{\mathcal G} \frac{\vec v_E^2}{2}+ \int_{\mathcal T} \frac{\vec v_B^2}{2}
\end{eqnarray}

The last term is an anomaly, calculated in \cite{M23PR}. Up to negligible power corrections in $R/R_C, w/R$
\begin{eqnarray}
    &&\int_{\mathcal T} \frac{\vec v_B^2}{2} = \frac{\Gamma_B^2}{8\pi}\oint_C d l \left(\gamma +\log \frac{c(l) R^2}{8 \nu} \right);\\
    && c(l) = \vec C'(l) \cdot \hat S\left(\vec C(l)\right) \cdot \vec C'(l) 
\end{eqnarray}

The term with $\log R$
\begin{eqnarray}
    \frac{\Gamma_B^2}{4\pi} \oint_C d l \log R
\end{eqnarray}
is canceled by a similar term coming from the Euler field. 
This cancellation is a consequence of the matching conditions, as discussed in \cite{M23PR}. Here are these calculations. 
Taking the derivative in $R$, we find the contribution of the surface $\partial \mathcal T$
\begin{eqnarray}
   && \vec v_E(\vec r) \ra \frac{\Gamma_B}{2 \pi} \frac{ \left(-y,x,0\right)}{x^2 + y^2};\\
   && \partial_R \int_{\mathcal G} \frac{\vec v_E^2}{2}  = -\int_{\partial \mathcal T} \frac{\vec v_E^2}{2} \ra -\frac{\Gamma_B^2}{8 \pi^2}\int_C d l  \frac{2 \pi}{R} = -\frac{\Gamma_B^2}{4 \pi } \oint_C d l \frac{1}{R}
\end{eqnarray}
which adds up to zero with the derivative of the above $\log R$ term.
Therefore, the sum of these two terms in the Hamiltonian does not depend on $R$ ( up to neglected power corrections).

As a result of this independence, we can take a limit $R \ra 0$ in the regularized Euler part and finally find the anomalous Hamiltonian
\begin{subequations}\label{AnomalousHamiltonian}
    \begin{eqnarray}
   && H = H_E + \frac{\Gamma_B^2}{8\pi} \oint_C d l \left(\gamma +\log \frac{c(l) |C|^2}{8 \nu} \right);\\
   && H_E = \lim_{R \ra 0}\left(\int_{\mathcal G} \frac{\vec v_E^2}{2}+ \frac{\Gamma_B^2}{8\pi} |C| \log\frac{R^2}{|C|^2}\right);
\end{eqnarray}
\end{subequations}

The flow $\vec v(\vec r)$ minimizing this Hamiltonian has the common flaw of all Euler flows (except topological ones): it continuously scales down to zero $\vec v_E =0, \Gamma_B =0$ by changing the scale of the velocity field.

Therefore, the zero velocity provides the absolute minimum of the Hamiltonian unless there are some topological restrictions on a flow, preventing continuous scaling to zero within a topological class.

\section{Topological Euler Flow}

\subsection{The Faddeev variables} 
The way to build topological solutions to the Euler equations was discovered in the eighties \cite{KM80, L81}.
These are spherical \CL{} or Faddeev variables, as we suggested in \cite{M23PR}.
These variables are elements of 2-sphere geometrically, 3D rigid rotators in the Hamiltonian dynamics, or the $\bS^2$ sigma model in the statistical field theory language.

The vorticity is locally parametrized as
\begin{eqnarray}\label{CLS2}
    &&\vec \omega = \oh Z e_{a b c}  S_a  \vec \nabla S_b \times \vec\nabla S_c;\\
    &&S_1^2 + S_2^2 + S_3^2= 1;
\end{eqnarray}
 where $Z$ is some parameter with the dimension of viscosity, staying finite when $\nu \ra 0$. 

 One could use various coordinates on the sphere. In particular, there are canonical coordinates
 \begin{eqnarray}
  &&\phi_1 =Z S_3;\\
  &&\phi_2 = \arg{ (S_1 + \I S_2)};
 \end{eqnarray}

 With these coordinates, the vorticity becomes parametrized as with the ordinary \CL{} variables
 \begin{eqnarray}
     \vec \omega = \vec \nabla \phi_2 \times \vec \nabla \phi_1
 \end{eqnarray}
 except these two variables $\phi_1, \phi_2$ vary on a rectangle $-Z \le \phi_1 \le Z, -\pi \le \phi_2 \le \pi$ rather than the whole plane $R_2$.

The Euler equations are then equivalent to passive convection of the \CL{} field by the velocity field (modulo gauge transformations, as we argue in \cite{M23PR}):
\begin{eqnarray}\label{CLEq}
     &&\partial_t \phi_a = -\vec v \cdot \vec \nabla \phi_a\\
     &&\vec v =\left(\phi_2 \vec \nabla \phi_1\right)^\perp; 
\end{eqnarray}
Here $V^\perp$ denotes projection to the transverse direction in Fourier space, or:

\begin{equation}
    V^\perp_\alpha(r) = V_\alpha(r) + \dal \dbe \int d^3 r' \frac{V_\beta(r')}{4 \pi |r-r'|}
\end{equation}

As we can see from this representation of velocity, it has a gap 
\begin{eqnarray}
    \Delta \vec v(\mathcal S) = 2 \pi n \vec \nabla \phi_1(\mathcal S)
\end{eqnarray}
at the surface where the phase $\phi_2$ has the gap $\Delta \phi_2 = 2 \pi n$.

\subsection{Boundary conditions at the surface and its edge}\label{boundaryConditions}

This surface is bounded by a singular line $C$ where the angular velocity in the transverse plane diverges as $1/r$, preserving a finite circulation for an infinitesimal dual loop $\delta C$ encircling $C$ (Fig.\ref{fig::DeltaC}).

There is only a tangent discontinuity of velocity at the surface $S_C\setminus C$,  coming from the delta function in vorticity. The boundary values are
 \begin{eqnarray}
     &&\vec \omega \ra 2 \pi n \delta(z) \vec \sigma \times \vec \nabla \phi_1;\\
     \label{velGap}
     && \vec v^+ - \vec v^- =  2 \pi n \vec \nabla \phi_1;\\
     && S_3^+  = S_3^-;\\
     && \phi_2^+ =\phi_2^- - 2 \pi n;\\
     && \omega^+_n =  \omega^-_n = - \vec \sigma \cdot \vec \nabla \phi_2^\pm \times\vec \nabla \phi_1^\pm
 \end{eqnarray}
 Here $\vec \sigma $ is the local normal vector to the surface, and $z$ is the normal coordinate.

The velocity gap must vanish at the edge of the surface, which requires the boundary condition
\begin{eqnarray}
    \vec \nabla S_3(C) =0;
\end{eqnarray}
On a simple surface with a boundary $\partial S = C$, as a consequence of incompressibility, the velocity  would be directed along the edge, i.e., the velocity $v_n$ along the inner normal to the loop will vanish
\begin{eqnarray}
    v_n = \vec v(\vec C(l)) \cdot \vec C''(l)/\abs{\vec C''(l)} \stackrel{?}{=} 0
\end{eqnarray}

However, our velocity is singular at the edge, and the Euler solution only applies outside a thin tube surrounding $C$. We must match this velocity with the Burgers solution at the surface of this tube (see Fig.\ref{fig::TubeSpiral}, Fig.\ref{fig::EulerOutside}).

It remains a computational problem to build a full velocity field, at least for the simple example of a flat circle $C$; here, we just sketched the flow using its general properties, which follow from our equations. These parametric 3D plots were created in \cite{MB19} with \Mathematica{} by solving the model particle motion equation in a velocity flow with the Kelvinon topology and geometry.

The singular rotational part of Euler velocity $\vec v_\theta = \frac{\Gamma_B}{2 \pi r}$ (matching the Burgers vortex inside) will contribute to the "normal inner" velocity $v_n$, at the junction with the tube; the result will depend on the angle $\beta$ in this plane normal to $\vec C'$ and diverge in the Euler limit.

Together, the rotational and axial flow \textbf{inside} the tube and inside the vortex sheet results in a spiral motion, as shown in Fig. \ref{fig::TubeSpiral}. 
In Fig. \ref{fig::TubeSpiral}, the thin spirals inside the green vortex sheet are tangent to both sides. Each spiral's big (alpha) cycle leads to vorticity normal to the vortex sheet, responsible for the circulation $\Gamma_C$. The small (beta) cycles of each spiral correspond to rotation around the local direction of the spiral, which corresponds to large tangent vorticity inside the vortex sheet.

In addition, there is an Euler flow outside the tube, which follows the Burgers spiral, but cannot penetrate the vortex sheet, as the normal velocity vanishes at both sides. 
Therefore, the rotational motion around the small cross-section of the torus does not go through the vortex sheet.

The external flow is sketched in Fig.\ref{fig::EulerOutside}. Vorticity points in a local direction of each spiral; the rotation velocity is orthogonal to vorticity, and the potential part of velocity is parallel to vorticity, providing helicity. The tornadoes are attached to the vortex sheet providing the spiral motion of particles towards the vortex sheet on both sides, separate from the spiral motion around the Burgers tube (see discussion below).

One can see that the circulations $\Gamma_C, \Gamma_B$ have opposite signs, in agreement with the Kelvinon theory. The tornadoes can only move towards the vortex sheet as the negative normal strain $S_{n n} <0$ leads to the exponential decay of the normal distance $z \propto\exp{t S_{n n}}$. Therefore, the tornadoes rotate clockwise, corresponding to negative circulation $\Gamma_C$, screwing the flow onto the surface in Fig.\ref{fig::EulerOutside}.

The velocity in the center of the spiral flow around the tube also goes clockwise to provide negative $\Gamma_C$, but this makes the circular motion in the cross-section plane go anticlockwise, meaning positive circulation $\Gamma_B$.
Note that the trajectories are tangent to the vortex sheet at the surface. Vorticity points in a local direction of each spiral; the rotation velocity is orthogonal to vorticity, and the potential part of velocity is parallel to vorticity, providing helicity. 

The rotation of tornadoes on both sides must be directed towards the sheet, pressing the flow particles to this sheet on both sides. With some thinking, it becomes clear that the spiral motion around the tube cannot flow into the tornadoes on both sides -- it is either one or another tornado that will rise particles from the surface instead of moving them to the surface as stability requires.

The only way out of this topological puzzle is the following scenario. The external spiral motion around the tube never connects to tornadoes; these tornadoes bring the flow from infinity to each side of the vortex sheet independently of the spirals around the tube. 

These spirals make a full circle around, passing through the inner side of the tube adjacent to the sheet. The normal to-the-sheet velocity does not vanish in this region; instead, there is still a rotational movement around the tube.

The cross-section of this flow is shown in Fig.\ref{fig::ExternalFlowCut}.

\pct{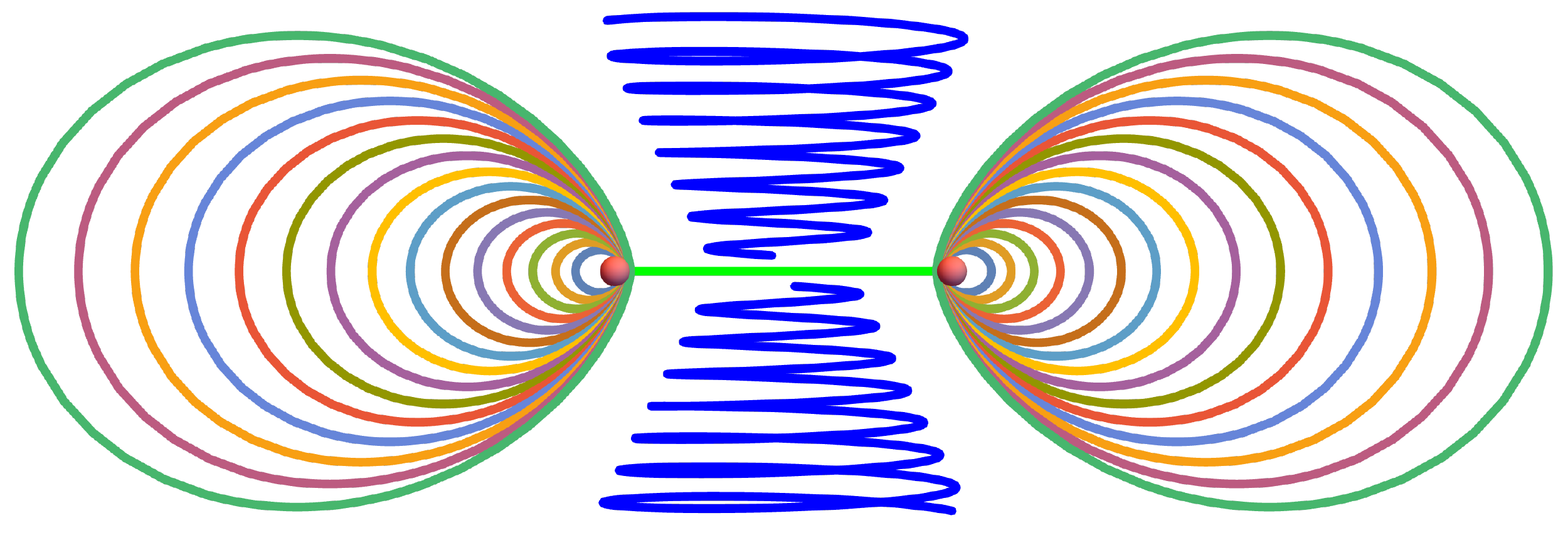}{The cross-section of external flow. The green horizontal line is a cross-section of the vortex sheet, the small red dots are the tube cross-sections, the ovals are the spiral particle trajectories projected on the $ x y$ plane, and the blue vertical spirals are tornadoes.}

\pct{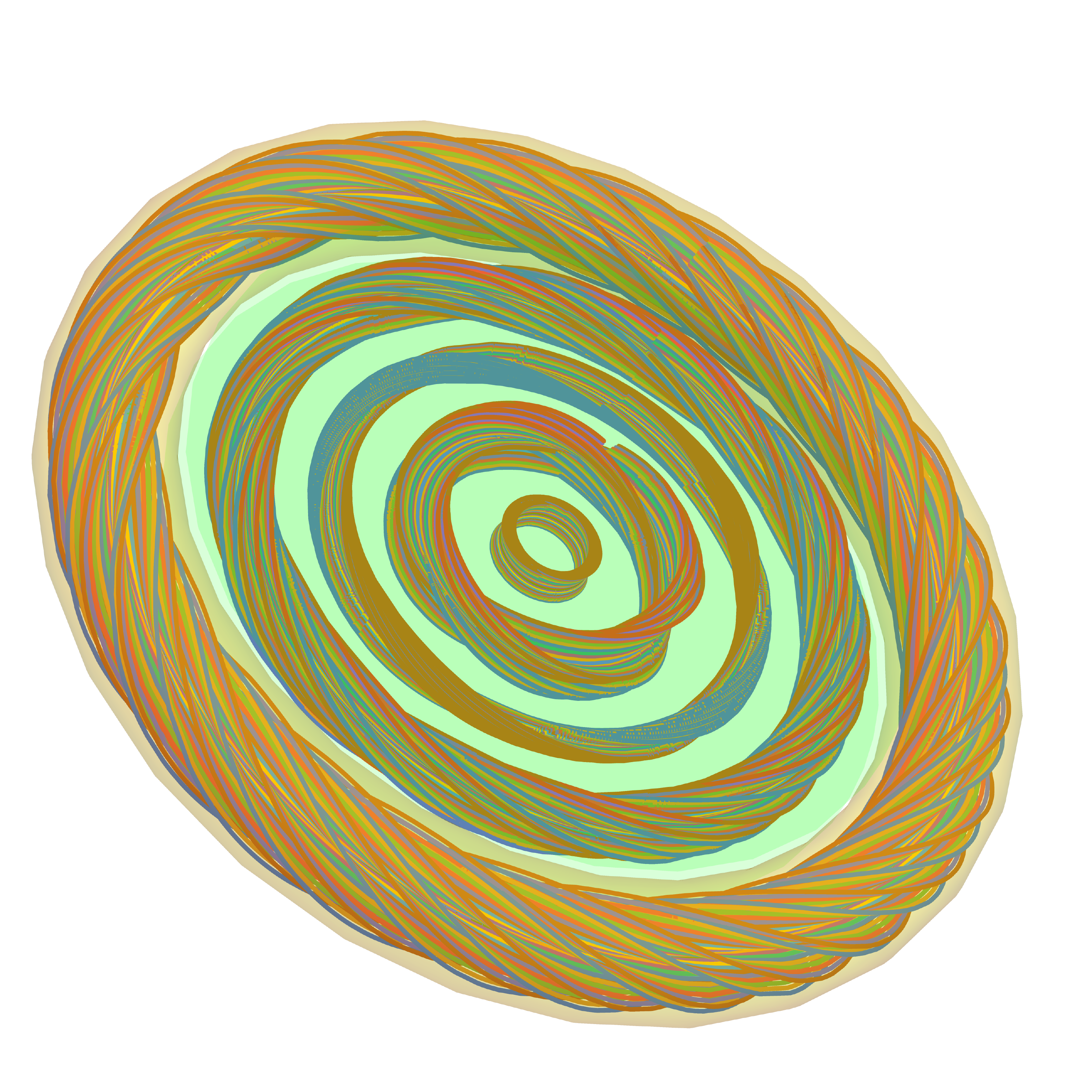}{\textbf{Inner \NS{} flow}. The idealized spiral motion of liquid particles around the axial vortex line \textbf{inside} the vortex tube (transparent yellow) and vortex sheet (transparent green). 
The thickness of the vortex sheet and the vortex tube is magnified compared to the turbulent regime we are studying. We trace 25 particles with different colors. }

\pct{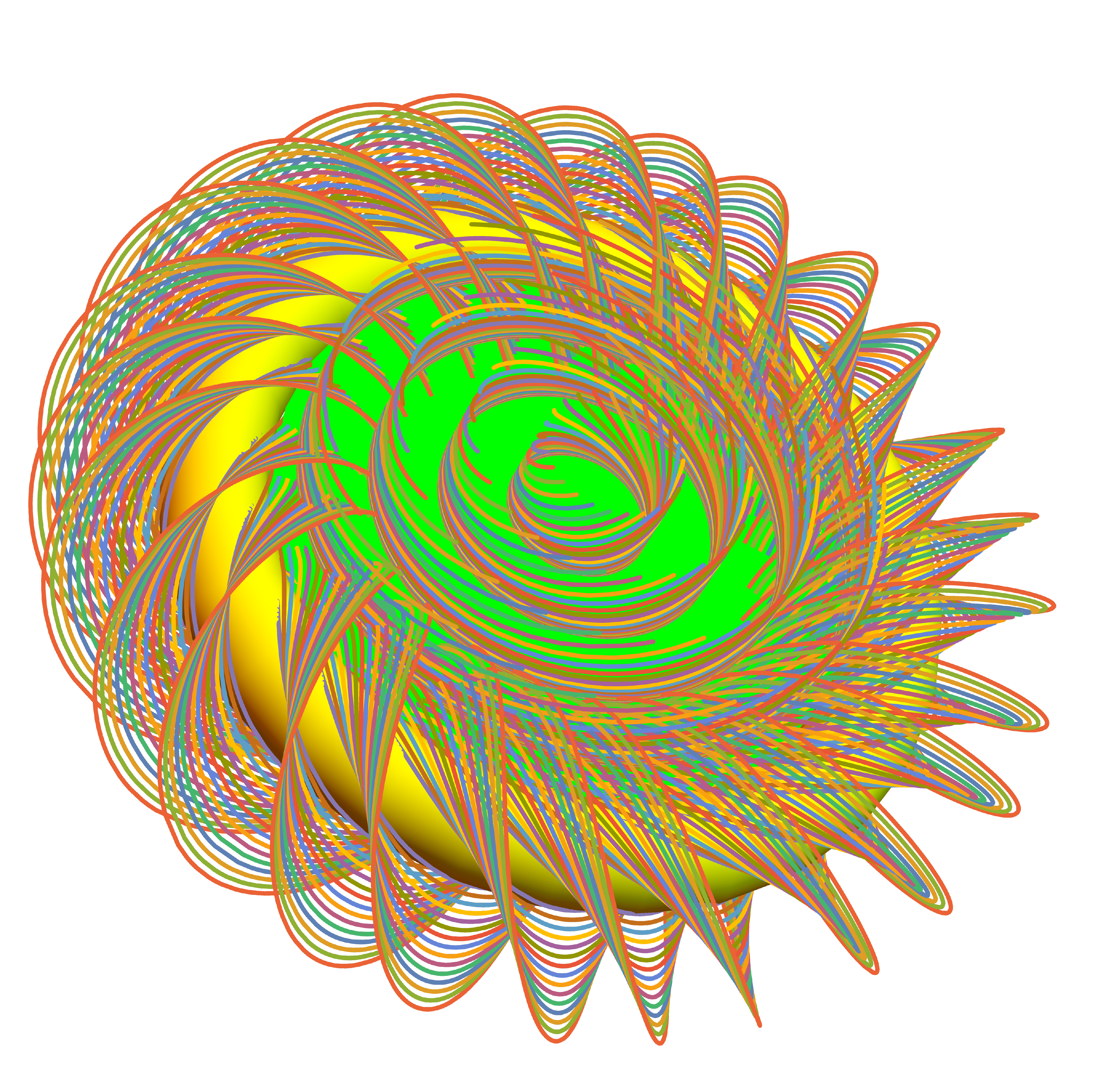}{\textbf{Outer Euler flow}. The idealized spiral motion of liquid particles around the axial vortex line \textbf{outside} the vortex tube (solid yellow)  with the tangent velocity boundary conditions at the vortex sheet (solid green). This flow goes around the tube, sliding along the surface on each side rather than passing through it.
The thickness of the vortex sheet and the vortex tube is magnified compared to the turbulent regime we are studying. We trace 25 particles with different colors. }

Let us note in passing that this singular tube and change of the Neumann boundary conditions at the edge invalidate the conditions of the de Lellis-Brue theorem in our case. 

This theorem \cite{M23PR}, in particular, claims that the surface Laplacian cannot be positive at the whole disk-like vortex sheet if the velocity is nonsingular and tangent to the edge.

We need this positivity for stability, as it was argued in \cite{M23PR} as part of the \CVS{} conditions.

Let us come back to our boundary conditions.

The potential velocity part enters the strain near the loop and has no singularity.

The true boundary condition is an eigenvalue requirement \eqref{EigenvalueCondition}. 

The analogous \CVS{} conditions \cite{M23PR} for the vortex sheet were equivalent to the velocity gap being a null vector of a boundary strain at the surface
\begin{eqnarray}\label{strainDotGap}
    \hat S\cdot\Delta \vec v =0.
\end{eqnarray}
The shape of the vortex sheet in \cite{M23PR} was fixed by this requirement, up to a few parameters.

We expect the same relation here to provide the same vortex sheet stability.

However, the velocity gap vanishes at the boundary, so this relation does not fix the boundary value of the strain. The eigenvalue requirement represents this missing condition at the edge of the discontinuity surface.

\textit{We conjecture that in the case of the Kelvinon, the shape of the discontinuity surface is determined by \eqref{strainDotGap} inside, plus \eqref{EigenvalueCondition} at the edge.}

Note that here we deviate from our old hypothesis \cite{M21a, M21b, M21c, M21d} that the shape of the discontinuity surface minimizes its area.

This minimal area would be an asymptotic solution of the loop equation at large smooth loops \cite{M23PR}, but not the exact shape of the discontinuity surface at an arbitrary loop.

Note also that as a consequence of \eqref{EigenvalueCondition}, the eigenvalues of a strain at the surface  have a form of
\begin{eqnarray}
    \diag{-\lambda(\vec r),0, \lambda(\vec r)}
\end{eqnarray}

The highest eigenvalue $\lambda(\vec r) >0$ has the eigenvector in the tangent plane of the surface, orthogonal to the velocity gap \eqref{velGap}, as it follows from \eqref{strainDotGap} and stability requirement (the normal component of strain at the surface must be negative).

Thus, if the velocity gap is orthogonal to the surface edge near the edge, the leading eigenvector of the strain would indeed be directed along the loop, as we required.

This condition is compatible with the fact that $\vec \nabla S_3$ vanishes linearly near the edge
\begin{eqnarray}
&& \vec \nabla S_3(\vec r \ra \vec C(l_0) )\propto (\vec r - \vec C(l_0)) ;\\
&&\Delta \vec v(\vec r \ra \vec C(l_0) ) = 2 \pi n Z \vec \nabla S_3 \propto (\vec r - \vec C(l_0)) \perp \vec C'(l_0);
\end{eqnarray}

The last relation follows from  the fact that $\vec C(l_0)$ is the nearest point at the edge to the point $\vec r \in S$, which leads to 
\begin{eqnarray}
    \pd l (\vec r - \vec C(l_0))^2 \propto \vec C'(l_0) \cdot (\vec r - \vec C(l_0)) =0
\end{eqnarray}

It would be very interesting to find the Kelvinon solution for the flat circular $C$, where the discontinuity surface is the unit disk $D^2$.

 \pct{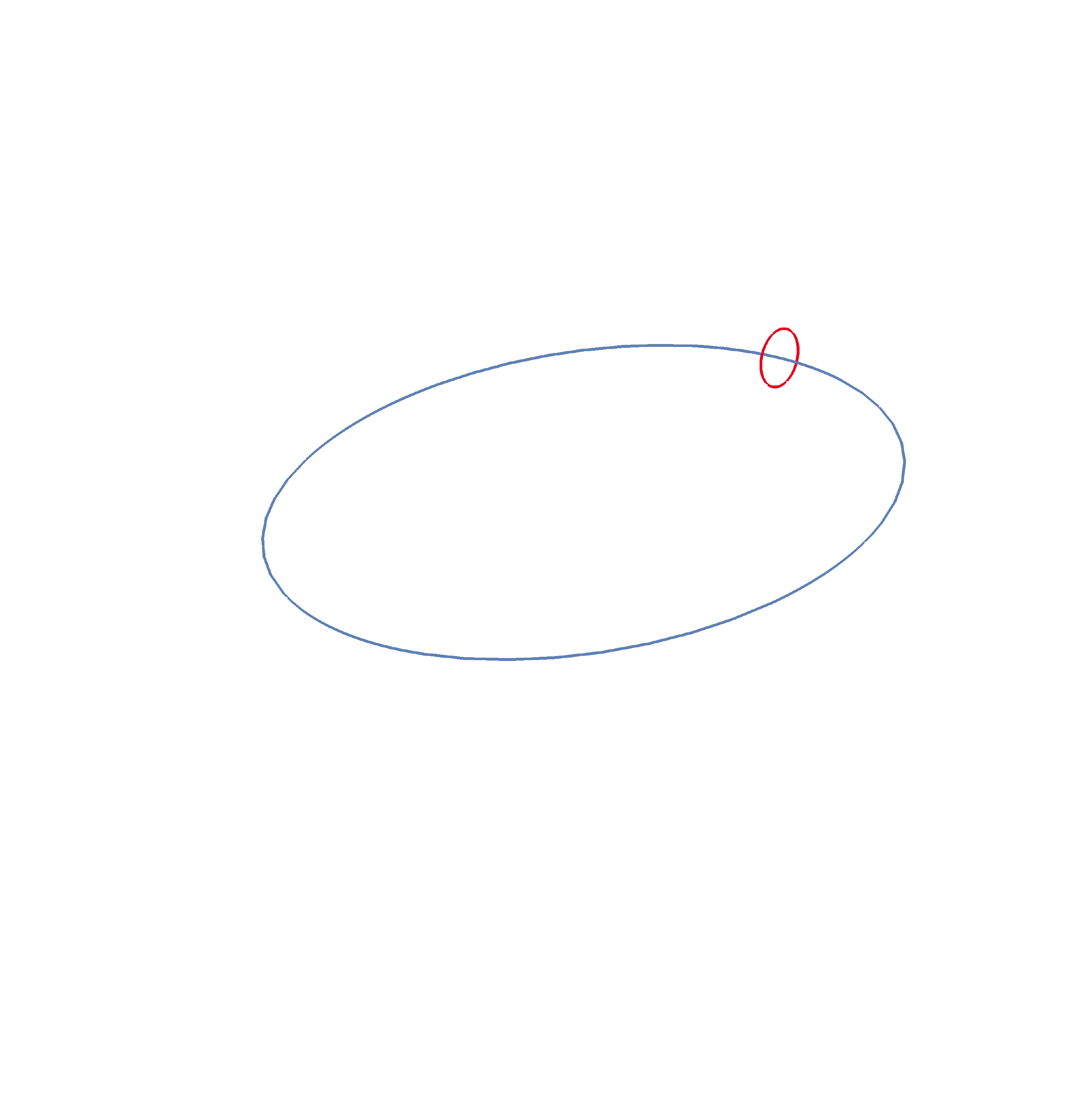}{The dual loop (red) encircling the monopole ring (blue). The Burgers vortex resolves the singular vortex line. }

\subsection{The mapping onto a disk on a sphere}

This tube cross-section's boundary $\delta C$ is mapped on the loop $\gamma$ on $\bS^2$, covered $n $ times.
Topologically, a circle is mapped on a circle with homotopy $\pi_1(\mathbb S_1) = \bZ$.

The circulation $\oint_{\delta C} \val d \ral$ is related to the vorticity flux through the tube's cross-section, which equates it to the Burgers parameter $\Gamma_B$.

The circulation around a loop on a 2-sphere equals the area of one of two complementary spherical caps $\Omega_\pm;\; \partial \Omega_\pm = \gamma$, depending upon the orientation of $\gamma$, i.e., the sign of the winding number $n$ (see Fig.\ref{fig::OmegaPM}).

\pct{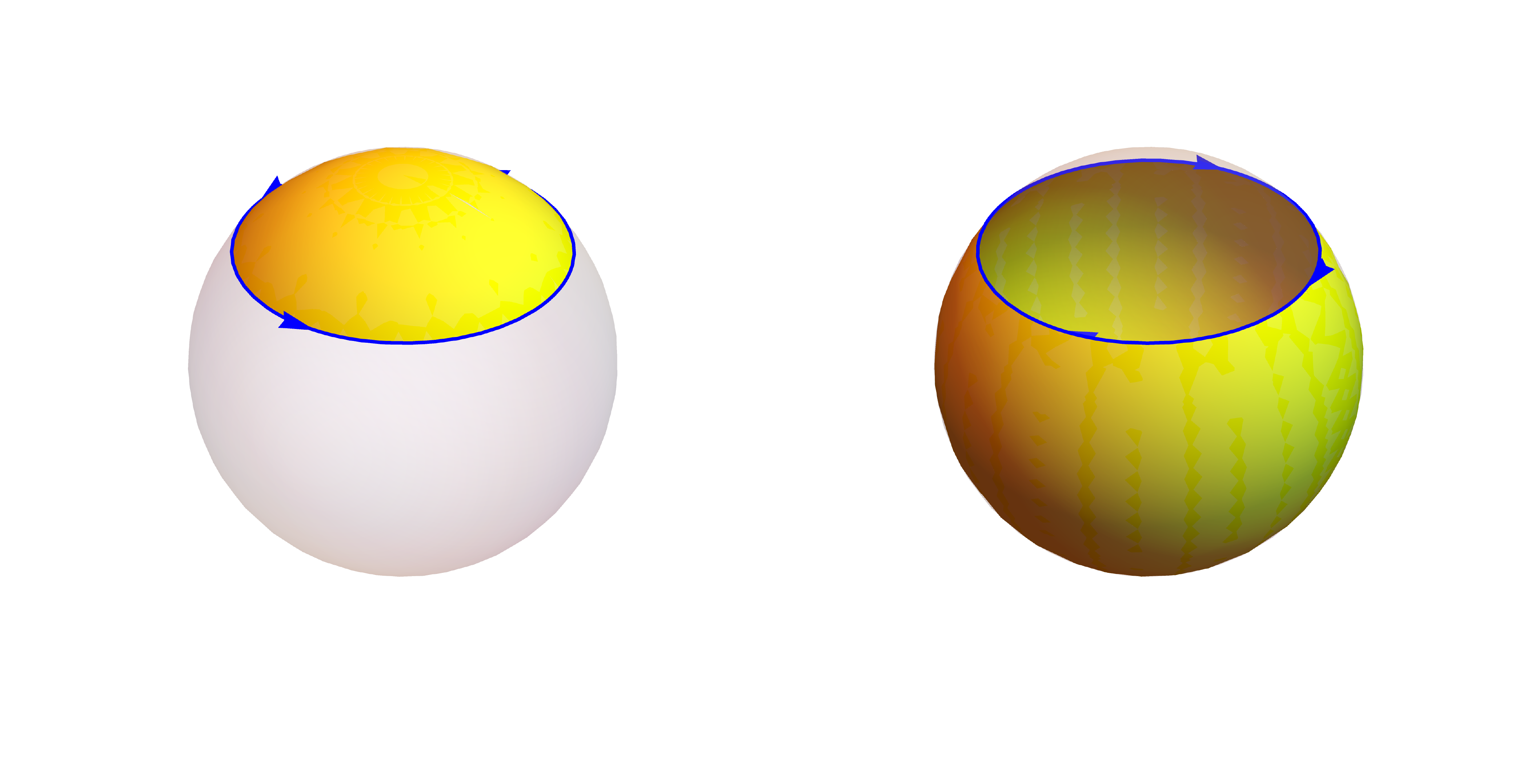}{The regions at $\Omega_\pm \in \bS^2$, with opposite orientations of the boundary loop $\partial \Omega_\pm = \gamma$. The areas $|\Omega_\pm|$ add up to $4 \pi$.}

In our coordinates, this loop $\gamma$ is some horizontal circle $S_3 = \cos\lambda =\const$.

\begin{eqnarray}\label{GammaB}
    &&\Gamma_B = \oint_{\delta C} \val d \ral  = \int_{\Omega_\pm} d \phi_2 \wedge d \phi_1=  2 \pi n Z(1 - \sign{n}  \cos\lambda);
\end{eqnarray}

The boundary value of $S_3(C)= \cos(\lambda)$ remains as a constant parameter of our \CL{} field, to be determined later from the minimization of the Hamiltonian.

There is, of course, a possibility to get zero circulation (and, therefore, zero velocity, in case $Z=0$ or $\cos \lambda = \sign n$).

Let us now turn to the circulation $\Gamma_C$ around the original loop. In the same way, as with the Burgers circulation $\Gamma_B$, this circulation can be written as a vorticity flux through some Stokes surface bounded by the loop and passing through the Euler region.

A small part of this surface will pass through the Burgers tube. As the vorticity inside the Burgers tube is directed towards its axis,  we can choose this surface to pass this tube in the local tangent plane to the loop. The normal to this local tangent plane $ \vec n \parallel \vec C' (l) \times \vec C" (l)$ is orthogonal to the direction of the vorticity $\vec \omega \parallel \vec C' (l)$.

Therefore, the flux is determined solely by the vorticity in the Euler region, which means that the singular-line solution \eqref{SingularLineVelocity} with potential flow outside the loop does not provide a finite circulation $\Gamma_C$.

This requirement is the ultimate reason for the topological solution.

There is no contradiction at this level with the \CL{} field, but there is an interesting relation based on the Stokes theorem.

Let us compute the Euler flux through the discontinuity surface on the upper side $\mathcal S^+_C$.

The flux through the surface is
\begin{eqnarray}\label{GammaC}
    &&\Gamma_C =  \int_{S^+_C} d \vec \sigma \cdot \vec \omega^+ =\int_{\Omega_\pm} d \phi_1 \wedge d \phi_2 = 2\pi m Z(1 - \sign{m} \cos\lambda);
\end{eqnarray}
where $m \in \bZ$ is a winding number for $\phi_2$ around the loop $C$.

These two winding numbers $n, m$ are consistent with the boundary condition \cite{M23PR} at the surface $\partial \mathcal T$ of the infinitesimal tube $\mathcal T$
\begin{subequations}
\begin{eqnarray}
\label{Phi2Map}
    &&\phi_2\left(\vec C(l) + \vec \xi\right) = m \alpha + n \beta;\\
    && \vec \xi =\eps \left( \vec n \cos\beta + \vec \sigma \sin\beta \right);\\
    &&\vec n = \frac{\vec C''}{|\vec C''|};\\
    &&\vec \sigma = \vec C' \times\vec n;\\
    && \alpha = \frac{2 \pi l}{|C|};\\
    && |C| = \oint_C d l;
\end{eqnarray}
\end{subequations}
Here, $\eps \ra 0$ is the radius of the tube.

\subsection{The topology}
Let us briefly discuss the topology of the Kelvinon.
The equation \eqref{Phi2Map} implements the mapping of a torus on a circle with homotopy $ \pi_1(\mathbb T^2) \cong \bZ \times \bZ$, which corresponds to a pair of integer winding numbers $n, m \in \bZ$.

What about mapping the 3D space by the \CL{} field $S_a(\vec r)$?

The Hopf mapping of the compactified 3-space to a 2-sphere,  $\bS^3 \mapsto \bS^2$, was already implemented by a spherical \CL{} field in \cite{KM80, L81}, but our Kelvinon is different.

The Kelvinon implements a mapping of the compactified 3-space $\mathcal G \cong\bS^3$ \textit{without a monopole ring} $\mathcal T \cong \bT^3$ onto the spherical cap $\Omega_\pm\cong \bD^2$ rather than the full sphere $\bS^2$.

Topologically,
\begin{eqnarray}
&& \mathcal G \cong  \bS^3 \setminus \bT^3 \cong \bT^3 ;\\
 &&  \bT^3\mapsto \bD^2;
\end{eqnarray}
As we have seen, this last mapping $\bT^3\mapsto \bD^2$ is described by two winding numbers. Depending upon the signs of these winding numbers, the \CL{} field maps physical space on one of the two complementary caps on a sphere separated by a circle $S_3 = \cos\lambda$.

The Appendix presents a family of smooth \CL{} fields with desired properties, including winding numbers $n,m$ and a decrease of vorticity at infinity. We do not compute velocity for these examples, as it would require a solution of a nontrivial Neumann problem on a minimal surface bounded by an arbitrary smooth loop $C$.



The Euler velocity field maps the compactified physical space $\bS^3$ \textbf{without the vortex surface} onto $\bR^3$. The vortex surface (with its edge $C$) is excluded because the velocity field has a gap at that surface. The space we excluded is a topological 3-disk $\bD^3$, and so is the remaining space.  

There are no topological invariants associated with this mapping by velocity field, but there are such invariants for the mapping  of the solid torus on a 2-disk by the spherical \CL{} field, as we discussed above.

Comparing the expression \eqref{GammaC} to Burger's contribution to the line integral of velocity at the center of the core of the tube, we get a self-consistency relation between parameters of the Euler flow
\begin{equation}\label{SelfConsistencyGammaC}
    \oint_C \ral   S_{\alpha\beta} d \rbe =-2\pi m Z(1 - \sign{m} \cos\lambda);
\end{equation}

This relation being linear, the normalization factor $Z$ can be canceled so that this is a restriction on a Faddeev vector field $S_a(\vec r)$.
The eigenvalue equation \eqref{EigenvalueCondition} also restricts the field $S_a(\vec r)$, leaving the normalization factor arbitrary.

As suggested in \cite{M23PR}, this factor will be determined from the energy balance between the incoming energy flow and anomalous dissipation.

There is some advance we have recently made in this part of the Kelvinon theory; we present it in the next Section.

\section{Energy balance revisited}

For every stationary solution of the \NS{} equation, the time derivative of any functional of the velocity field, including the Euler Hamiltonian, must vanish.

Our matching principle suggests using this \NS{} relation to fix the remaining free parameter $Z$ of the Euler flow in the \CL{} variables.

The energy dissipation is localized in the Burgers vortex core. We know it in an inviscid limit as a functional of the local Euler strain at the loop \eqref{AnomalousDissipation}. At fixed \CL{} field $S_a(\vec r)$, this dissipation is proportional to $Z^3$ (two powers of $Z$ from circulation $\Gamma_B$ and one from the strain.)

The matching energy pumping into this region can be written in many forms. In the review paper \cite{M23PR}, we used the energy flux through the boundary of the volume $V$, which flux we then estimated as $\eps V$.

There is a more direct approach, using notorious external random forces $\vec f(\vec r)$, which we take as a random uniform vector inside this volume.

This point needs clarification. Usual forces are also a function of time, correlated by a $\delta(t-t')f(\vec r - \vec r')$ with some slow function of space distance $\vec r - \vec r'$. Then the time averaging is assumed, leading to some expressions for the equal time velocity correlations.

In our approach, we replace time averages with ensemble averages. We have an ensemble of stationary flows, each with its uniform force $\vec f$ drawn from a Gaussian distribution. 
Our force is a space-independent Gaussian random vector with Kronecker delta variance
\begin{eqnarray}
    \VEV{f_\alpha f_\beta} = \sigma \delta_{\alpha\beta}
\end{eqnarray}

This ensemble averaging is equivalent to time averaging with delta-correlated Gaussian force (such force is a different sample of a Gaussian vector at different times).

The energy pumping created by such a force inside a volume $V$ before Gaussian averaging is simply the work made by this force over the net momentum of the fluid inside this volume
\label{epsEq}
\begin{eqnarray}
    &&\et =  \int_V \delta \vec v \cdot \vec f;
\end{eqnarray}

This perturbed velocity $\delta \vec v$, in turn, depends on the force through the \NS{} equation.
We shall assume this relation to be linear, and later we justify that assumption in the inviscid limit utilizing asymptotic freedom.

In the linear approximation, we have to solve linearized Euler equations for $\delta \vec v, \delta p$
\begin{subequations}\label{LinEq}
    \begin{eqnarray}
    &&\hat G\cdot \delta \vec v  +  \vec \nabla \delta p = \vec f;\\
    && \hat G_{\alpha\beta} = (\dbe \val) \hat I + \delta_{\alpha \beta} \vga \pd \gamma \;\\
    && \vec \nabla^2 \delta p = - \vec \nabla \cdot  \hat G\cdot \delta \vec v ;
\end{eqnarray}
\end{subequations}

Symbolically, we can write the result
\begin{eqnarray}\label{Qmatrix}
    &&\et = \vec f \cdot \hat Q\cdot \vec f;\\
    && \hat Q = \IINT V d^3 \vec r d^3 \vec r'\left(\hat G - \frac{1}{\vec \nabla^2} \vec \nabla \otimes \vec \nabla \cdot \hat G\right)^{-1}(\vec r, \vec r')
\end{eqnarray}
Let us count the factors of $Z$ here, assuming $\vec f$ to be $\MO 1$. As we have shown in \cite{M23PR} (see also the next Section), in the inviscid limit, there is asymptotic freedom:
$Z$ grows as a power of the logarithm of the effective Reynolds number 
\begin{eqnarray}
   && Z \sim \left(\log \RE\right)^\ot \ra \infty;\\
   && \RE = \frac{\VEV{c}_C |C|^2}{ \nu}
\end{eqnarray} 
Then, we have $\hat G \sim Z, \hat Q \sim 1/Z$ in the \eqref{Qmatrix}.

This makes the correction $\delta \vec v \sim 1/Z$ much smaller than $\vec v \sim Z$
\begin{equation}
    \delta \vec v /\vec v \sim Z^{-2} \sim \left(\log \RE\right)^{-\tt}
\end{equation}

Now we have justified linear approximation for the perturbation of the Euler equation by an external pumping force.

In general, asymptotic freedom in Turbulence makes the fluctuations of the velocity field around Kelvinon go to zero as a power of the logarithm of the Reynolds number. 

The same thing happens in QCD with fluctuations of the gluon field around the instanton: these fluctuations logarithmically die out compared to the instanton field.

As the background velocity fluctuation $\vec v_0 =\VEV{\delta \vec v}_V$ is a linear function of $\vec f$, the Gaussian distribution of $\vec f$ is equivalent to the Gaussian distribution of $\vec v_0$, which was considered in \cite{M23PR}. The difference is that we now have a microscopic equation \eqref{LinEq}, which allows us to compute this background velocity field once the base Kelvinon flow is known.

Let us turn back to the energy balance. Naturally, only the total energy of the \NS{} flow is stationary. There are some contributions $\ et'_d, \ et'_p$ from the remainder of the fluid  both to the energy pumping $\et_p$ and to the energy dissipation $\et_d$ in the energy balance equations
\begin{eqnarray}
     \et'_p +\vec f \cdot \hat Q \cdot \vec f - \sigma \tr \tilde Q=   \et'_d + \frac{\Gamma_B^2}{8 \pi}\oint_C d l c(l);\\
\end{eqnarray}
where we subtracted the mean over the random forces from the $\tilde Q$ term to satisfy the mean energy balance. The outside volume energy flow components $\ et'_d, \ et'_p$ are treated as constants rather than random numbers (self-averaging of the random forces acting in the remaining infinite volume.)

We have to factor our $Z$ and solve the resulting equation (with variables $\tilde X $ corresponding to  $X$ with $Z=1$)
\begin{subequations}
    \begin{eqnarray}
    &&Z^3 A = B + \frac{1}{Z}\left(\vec f \cdot \tilde Q \cdot \vec f  - \sigma \tr \tilde Q\right);\\
    \label{Aconst}
    && A = \frac{\tilde \Gamma_B^2}{8 \pi}\oint_C d l \tilde c(l);\\
    && B =  \et'_p - \et'_d ;
\end{eqnarray}
\end{subequations}

The unknown parameter $B$ here can be estimated as excessive energy pumped into our volume to be dissipated inside the singular vorticity tube.
This parameter is proportional to the missing volume in each of the energy flows $\ et'_d, \ et'_p$
\begin{eqnarray}
    B \approx \eps V[C] \propto \eps |C|^3
\end{eqnarray}
where $\eps$ is a \KO{} energy flow per unit volume and $V[C]$ is the volume occupied by our soliton around the singular loop $C$. 

The last estimate $V \propto |C|^3$ implies that the loop has only one scale, which can be taken as its length $C$. 

There are two scales for a large, almost flat loop with small normal deviations: the minimal area $|S[C]|$ and the width $\Delta$ of vorticity field distribution around this surface.
In this case $V[C] \sim |S[C]| |\Delta|$.

We postpone the discussion of large flat loops to the next section.

The solution of this quartic equation at small $A $ equals to
\begin{subequations}
    \begin{eqnarray}
   && Z \ra \left(\frac{B}{A}\right)^\ot  + \vec f \cdot \hat M\cdot \vec f - \sigma \tr \hat M;\\
   && \hat M = \frac{\tilde Q}{3 B} ;\\
   && A \sim \frac{1}{\log \RE} \ra 0
\end{eqnarray}
\end{subequations}

In the scaling region (where there is only one scale $\eps$), the fluctuating part of $Z$,  $\vec f \cdot \hat M\cdot \vec f$  must scale the same way as the first term in $Z$. Simple dimensional analysis shows that for that purpose, we need the following scaling law
\begin{eqnarray}
   \sigma \sim Z B/\tilde Q \sim \eps^{\ot} |C|^{\ft} \eps |C|^3 |C|^{-2} \sim \eps^{\ft} |C|^{\frac{7}{3}}
\end{eqnarray}

How can the variance of random forces be related to the size of the volume taken by the Kelvinon? 

These are not the symbolic forces at the boundaries of an infinite volume but rather effective forces acting on the boundary of this volume, representing the effects of random pressure terms coming from remote vortex structures.

This estimate must hold as long as we accept the K41 dimensional analysis for small enough loops compared to the intrinsic size of the Kelvinon.

Finding a microscopic mechanism leading to such self-consistent forces would be necessary.
\section{Asymptotic freedom revisited}

Now we can revisit and extend the analysis of \cite{M23PR} of asymptotic freedom (logarithmic decrease of running coupling constant of the Kelvinon theory with the local Reynolds number).

Let us elaborate on the relation \eqref{Aconst}. Using the basic formula \eqref{GammaB}, we find
\begin{subequations}\label{TildeConds}
    \begin{eqnarray}
    &&A = g\oint_C d l \tilde c(l) ;\\
    && g = \frac{ \pi n^2 (1- \sign{n}  \cos\lambda)^2 }{2} ;\\
    \label{S3Cond}
    && S_3(\vec C(l)) = \cos\lambda ;\ \forall l;\\
    \label{streainCond}
    && \tilde S_{\alpha\beta}(\vec C(l)) C'_\beta(l) = \tilde c(l) C'_\alpha(l);\\
    && \oint \ral d\rbe \tilde S_{\alpha\beta} = -2 \pi m ( 1 - \sign{m} \cos\lambda);\\
    && \tilde S_{\alpha\beta}(\vec r) = \frac{\dal \tilde \vbe + \dbe \tilde \val}{2};\\
    && \tilde \val = \left(\phi_2 \dal S_3 \right)_\perp;
\end{eqnarray}
\end{subequations}
The variables with tilde $\tilde S,\tilde v,\dots$ correspond to the normalized flow, with $Z \Ra 1$.

Factoring out the $Z-$ dependence in the anomalous Euler Hamiltonian \eqref{AnomalousHamiltonian}, we find
\begin{subequations}\label{TildeHamiltonian}
\begin{eqnarray}
     && H = Z^2 \left(\tilde H + g |C| \log\frac{Z}{\nu}\right) ;\\
   && \tilde H = g\oint_C d l \left(\gamma +\log \frac{\tilde c(l) |C|^2 }{8 }\right) + \lim_{R \ra 0}\left(\int_{\mathcal P} \frac{\tilde v_E^2}{2}+ g |C| \log\frac{R^2}{|C|^2}\right);\\
   && Z = \left(\frac{\eps V[C]}{\oint_C d l \tilde c(l) }\right)^\ot g^{-\ot}
\end{eqnarray}
\end{subequations}

This factor $g$ depends on only one dynamical variable: the boundary value $S_3(C) = \cos\lambda$.
The \CL{} fields $\phi_2, S_3 $ outside the loop $C$ satisfy the passive advection equations \eqref{CLEq}.
The boundary value of the strain $\tilde S$ satisfies the boundary conditions \eqref{TildeConds}.

The parameter $g$ must minimize the Hamiltonian, with fixed loop $C$ and fixed \KO{} constant $\eps$.

Varying the Hamiltonian \eqref{TildeHamiltonian} with respect to $g$, we find the transcendental equation for $g,Z$
\begin{eqnarray}
&& g = \frac{g_0}{-1 +\log \frac{Z}{\nu }};\\
&& g_0 = \frac{2 \tilde H}{|C|};\\
&& \left(-1 +\log \frac{Z}{\nu }\right)Z^3 = \eps R^4[C];\\
&& R^4[C] =\frac{g_0 V[C] }{\oint_C d l \tilde c(l) }
\end{eqnarray}

This $R[C]$ is a functional of the loop $C$. By dimensional counting, it scales as $|C|$
\begin{eqnarray}
    R[C] =   |C|f\left[ \frac{C}{|C|}\right]
\end{eqnarray}
This estimate assumes that there are no other scales in the solution. As mentioned above, for the smooth, almost flat loop $C$, there may be an effective scale in the normal direction to the minimal surface.

Neglecting such a case (it will be considered in the next Section), we can take $R=R[C]$ as a definition of the loop scale and write down the "renormalization group" equation for effective coupling $g_R =\frac{g}{g_0}$
\begin{eqnarray}
   && \dbyd{g_R}{\log R} = \beta(g_R);\\
   && \beta(g_R) = - \frac{4g_R^2}{3 + g_R};
\end{eqnarray}

This beta function corresponds to the transcendental equation for $g_R$
\begin{eqnarray}\label{eqforgR}
    \frac{3}{g_R} - \log g_R = -3 + 4 \log \left(\frac{\eps^\oq R}{ \nu^\tq}\right)
\end{eqnarray}

\pct{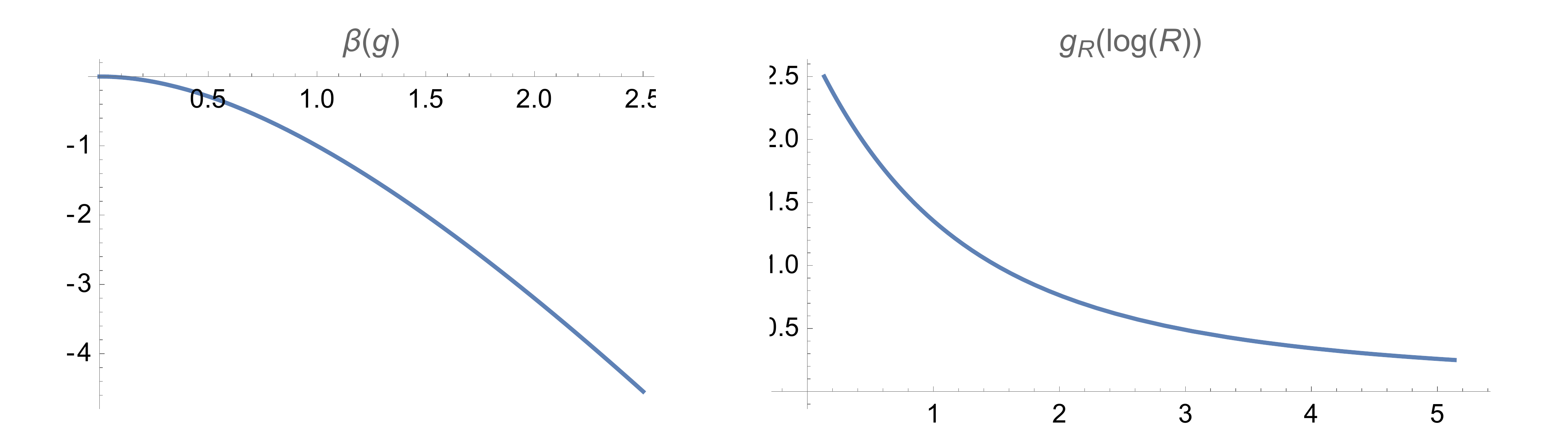}{The beta function $\beta(g)$ and the running coupling constant $g_R(\log R)$}

The beta function and the solution for $g_R(\log R)$ are plotted in Fig.\ref{fig::RunningCoupling}. This beta function has no positive roots; it monotonously goes to $-\infty$. 

The roots of the beta function would correspond to the fixed point of RG, leading to scaling laws with anomalous dimensions. \textbf{That} would justify the traditional multifractal scaling laws. However, in our theory, the beta function is calculable in explicit form and does not have a root.

Therefore, we have asymptotic freedom instead of the usual multifractal scaling laws. As discussed below, the difference is hard to observe in DNS and even harder in real experiments on Earth.

Now, with asymptotic freedom, we can justify the linearization of the Euler equation in the presence of fixed external random force: this is the leading term of expansion in our running coupling constant.

In the same leading approximation, the $Z$ factor grows as a power of $\log \RE$
\begin{eqnarray}
     Z \ra \left(\frac{\eps}{3} R^4\log \left(\frac{\eps R^4}{ \nu^3}\right)\right)^\ot 
\end{eqnarray}

\subsection{Anomalous Hamiltonian and the Matching Principle}

This approach to the stationary solutions of the \NS{} and Euler equation is unusual and may be confusing.
Let us summarize it and clarify the steps involved.

We use conventional Hamiltonian $H =\int_{R^3} \vec v^2/2 $ and conventional NS equations in whole space. The viscosity is neglected outside the singular regions, and the Euler solution is used in that outside region.
The stationary solutions of the Euler and NS equations are matched in the region of distances r from the singular line $w \ll r \ll R_C$, where $w$ is the local width of the Burgers vortex, and $R_C$ is the local radius of curvature of the loop.
This combined solution has some free parameters left, namely the normalization constant $Z$ for the Clebsch field in the outside region and $S_3(C) = cos \lambda$ is the boundary value of the Clebsch field component $S_3$ at the loop. 

It also depends on the shape of the discontinuity surface.
The parameter $Z$ is determined from the energy conservation on the NS stationary solution $\partial_t H = - E_d + E_p =0$
The shape of the surface is determined by the eigenvalue conditions on the surface and at the loop, which are part of the matching conditions at the surface of the tube.

We are left with the last parameter, $S_3$, to be determined from the minimum of the Hamiltonian.

\section{Multifractals}

For historical reasons, the multifractals are assumed to be pure scaling laws for the moments of the velocity differences, or circulation, in our case, as a function of the scale $R$.

Parisi and Frisch \cite{FP85}  borrowed this idea from Conformal Field Theory, where the moments of similar structure functions are described by scaling laws with anomalous dimensions. 

However, this was not the most general multifractal phenomenon in statistical field theory.

In the conformal field theory of critical phenomena in statistical mechanics, the anomalous dimensions are universal functions of the normal dimension of the fluctuating variable  (power of circulation in our case).

At the same time, in asymptotically free theories, like  QCD, the anomalous dimensions depend on the so-called running coupling constant, which tends to zero inversely proportional to the logarithm of scale.

So, the multifractal as a critical phenomenon has some precedent in statistical and quantum field theory. However, it is a dynamic question of whether the effective coupling is a universal number (a simple root of the beta function at finite coupling constant in the case of conformal theory) or it is running to zero like in asymptotically free theories (the double root of the beta function at zero coupling constant).

We can now answer that question for the Kelvinon theory in favor of asymptotic freedom \cite{M23PR}.

Our asymptotic freedom does not correspond to small velocity. On the contrary, velocity is large while its fluctuations are small; the same happens with the gluon field around instanton in QCD or with the electroweak field near the 't Hooft-Polyakov monopole (more relevant analogy, as we also have a monopole here).

The novel phenomenon in Turbulence is that the inverse powers of the Reynolds number do not appear in our expansion. In the inviscid limit, such terms become negligible, and the convergence to this limit slows down to the inverse powers of the logarithm of the Reynolds number.

This expansion in the zeroth approximation involves the nontrivial Kelvinon field $S_a(\vec r) \in \bS^2$, mapping the physical space (with added infinity and removed loop $C$) onto the spherical cap $\Omega_\pm \in \bS^2$.

There are two possibilities for the orientation $\sign n = \pm 1$, corresponding to an upper or lower cap on a sphere. These two possibilities are equivalent as they correspond to the reflection $S_3 \Ra -S_3$. In the asymptotically free limit, the boundary value $S_3(C) = \cos\lambda$ tends to the North or South pole, depending upon the sign of $n$.
\begin{eqnarray}
   \cos \lambda =  \sign n + \MO{(\log \RE)^{-\oh}}
\end{eqnarray}

This asymptotic formula will be compatible with our self-consistency relation \eqref{SelfConsistencyGammaC} for the circulation $\Gamma_C$ only if $\sign m = - \sign n$.
Then both sides of this equation are $\MO 1$ when the running coupling constant $g_R$ goes to zero.
\begin{eqnarray}
     && \oint_C \ral \tilde S_{\alpha\beta} d \rbe =-2\pi m (1 + \sign{n} \cos\lambda) \ra -4 \pi m ;
\end{eqnarray}
In this case, with the $\vec f$ correction in the linear approximation:
\begin{subequations}
    \begin{eqnarray}
    && \Gamma_C = 4 \pi m Z \left(1- \sqrt{\frac{g}{2 \pi n^2}}\right)\nonumber\\
    &&\ra  4 \pi m \eps^\ot R^\ft\left(g_R^{-\ot}- g_R^{\frac{1}{6}}\sqrt{\frac{g_0}{2 \pi n^2}}\right) + 4 \pi m \left(\vec f \cdot \hat M\cdot \vec f - \sigma \tr \hat M\right)
\end{eqnarray}
\end{subequations}

This expression for the velocity circulation has the same structure as the one in \cite{M23PR}
\begin{subequations}\label{GammaFromForce}
\begin{eqnarray}
    &&\Gamma_C = \tau + \vec \xi \cdot\hat q \cdot \vec \xi ; \\
    &&\vec \xi  \sim \mathcal N(0,1);\\
    && \tau = 4 \pi m \eps^\ot R^\ft\left(g_R^{-\ot}- g_R^{\frac{1}{6}}\sqrt{\frac{g_0}{2 \pi n^2}}\right) - 4 \pi m \sigma \tr \hat M;  \\
    && \hat q = 4 \pi m \sigma \hat M;
\end{eqnarray}
\end{subequations}

The non-fluctuating part $\tau$ of circulation shows the logarithmic growth with the scale $R$ through the running coupling $g_R^{-\ot} \sim (\log R)^\ot$, but the parabolic term $\vec \xi \cdot\hat q \cdot \xi$ does not depend on the running coupling constant. 

There are also higher-order terms of expansion in effective coupling $g_R$, which we have neglected.

According to \cite{M23PR}, this corresponds to the circulation PDF
\begin{eqnarray}
    W(\Gamma) \propto \frac{1}{\sqrt{|\Gamma -\tau|}} \exp{ -\frac{|\Gamma - \tau|}{q_0}}
\end{eqnarray}
where $q_0>0$ is the leading eigenvalue of the $3\times 3$ matrix $\hat q$.

The effective fractal dimension for higher moments, corresponding to the saddle point in the integral for the moments
\begin{eqnarray}
    &&\VEV{\Gamma^p} \propto \int d \Gamma  \Gamma^p W(\Gamma) \ra \textit{saddle point};\\
    && \lambda(p, \log R) \equiv \pbyp{\log \VEV{\Gamma^p}}{\log R} \ra  p \pbyp{\log q_0}{\log R} + \left(1 + \frac{1}{2 p}\right) \pbyp{(\tau/q_0)}{\log R}; \text{ if } p > p_c;
\end{eqnarray}

If there is no internal scale in a Kelvinon, the parameter $R$ is proportional to the size of the loop $|C|$, and the area inside the loop scales as $R^2$.
In that region, we have a K41 formula for $\lambda$, up to higher corrections in asymptotically free coupling constant $g_R$
\begin{eqnarray}
    \lambda(p, \log R) \ra  p \pbyp{\log \tau}{\log R} \ra \frac{4 p}{3}\text{ if } p < p_c;
\end{eqnarray}
As it was shown in \cite{M23PR} from an alternative dynamical theory (loop equations), the asymptotic dependence of the circulation PDF in the true inertial range $|\Gamma| \gg \nu, \eps^\ot|R|^\ft \gg \nu  $ must tend to the scaling law $ \Gamma \sim \sqrt{|S_{min}[C]|}$, where $|S_{min}[C]|$ is the minimal area of the surface bounded by $C$.

In our context, this means that $R^{\ft}$ is to be replaced by  $\sqrt{|S_{min}|}$. 

As for the next correction in $\lambda(p, \log \sqrt{|S_{min}|})$ at large $p$, it depends upon the dimensionless ratio $\tau/q_0$.

Asymptotic freedom tells us this ratio goes to infinity as in \eqref{GammaFromForce}. Differentiating that ratio and using the RG equation, we find
\begin{eqnarray}
   && \frac{\tau}{q_0} \ra  \const{}  \left(g_R^{-\ot}- g_R^{\frac{1}{6}}\sqrt{\frac{g_0}{2 \pi n^2}}\right) ;\\
   && \pbyp{(\tau/q_0)}{\log R} \ra  \const{} g_R^\tt \left(1 + \sqrt{\frac{g_0 g_R}{8 \pi n^2}}\right)
\end{eqnarray}
Thus, our prediction for the asymptotic behavior of the running index $\lambda(p, \log \sqrt{|S_{min}|})$ is
\begin{eqnarray}\label{lambdaPred}
    &&\lambda(p > p_c,\log \sqrt{|S_{min}|}) \ra  p + \const \left(1 + \frac{1}{2 p}\right) g_R^\tt \left(1 + \sqrt{\frac{g_0 g_R}{8 \pi n^2}}\right);
\end{eqnarray}
where $g_R$ is related to $\log \sqrt{|S_{min}|}$ by a transcendental equation \eqref{eqforgR}.
\pct{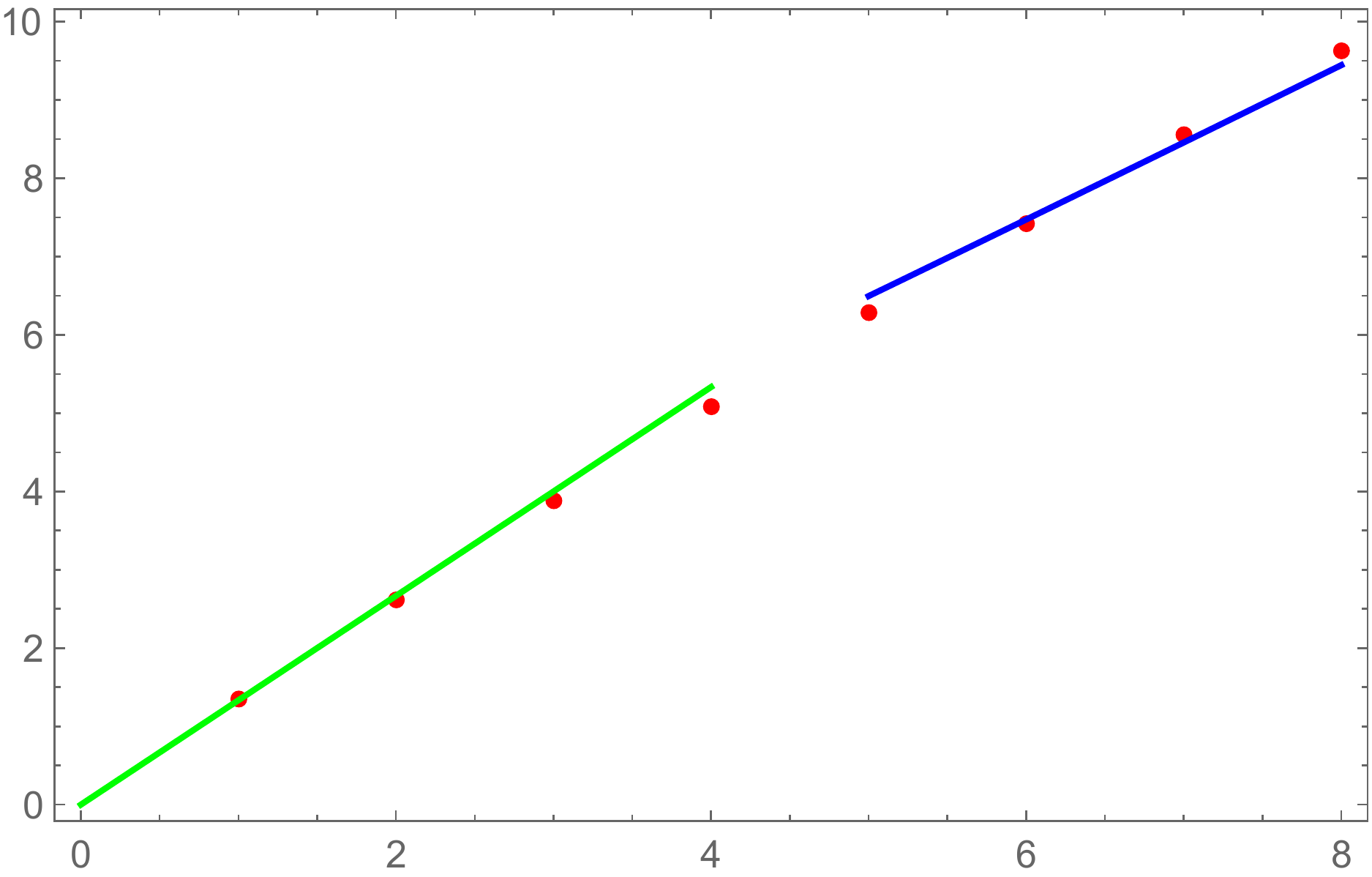}{DNS data \cite{S21} for classical circulation fractal dimensions $\lambda(p)$ fitted against our asymptotic formula \eqref{lambdaPred} (blue) and K41 line $\frac{4 p}{3}$ (green).}
\pct{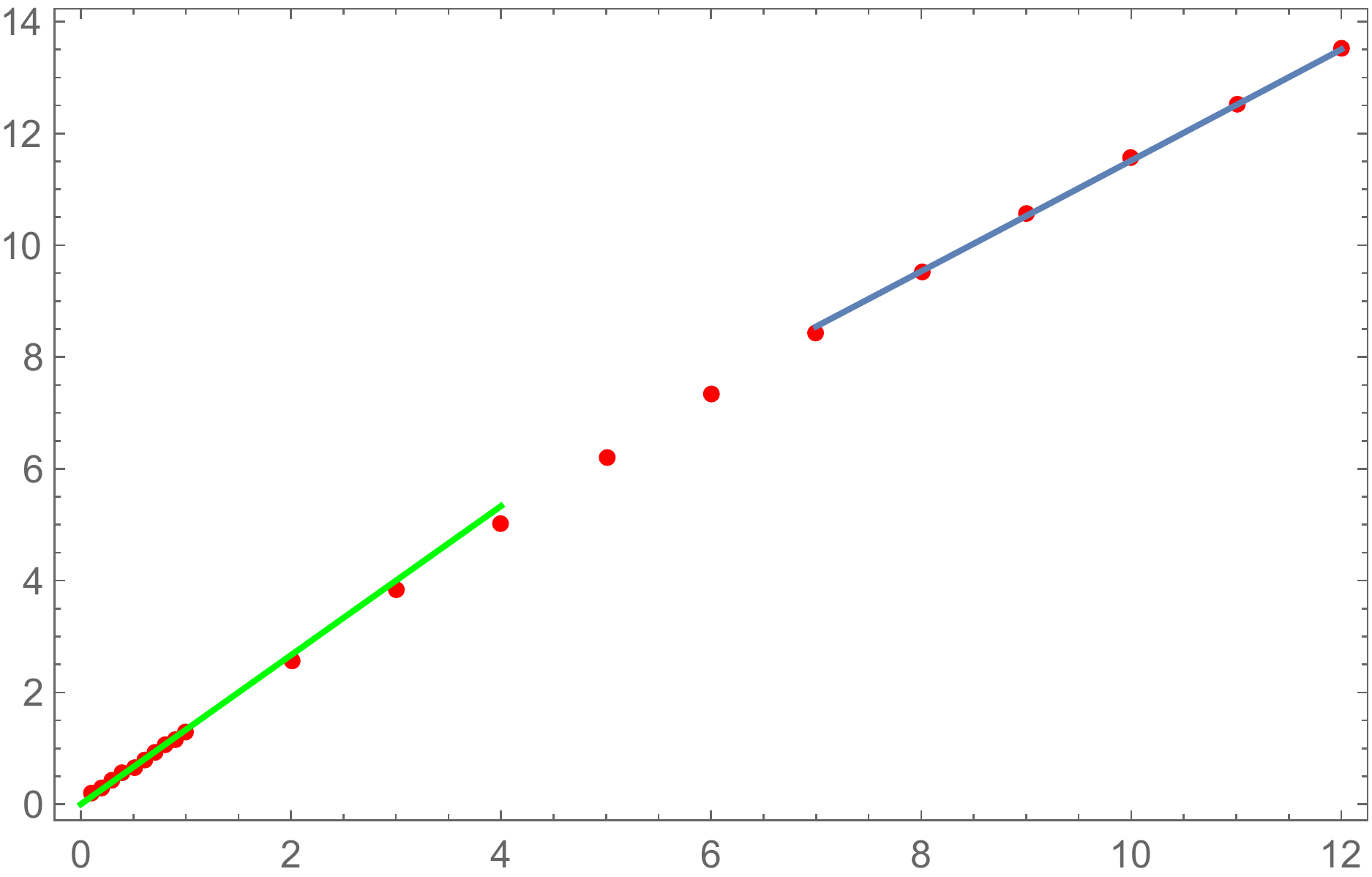}{DNS data  \cite{QuantumCirculation21} for quantum circulation fractal dimensions $\lambda(p)$ fitted against our asymptotic formula \eqref{lambdaPred} (blue) and K41 line $\frac{4 p}{3}$ (green).}

We do not have enough data to compare this formula for varying $\log \sqrt{|S_{min}|}$. However, we have the data obtained for the fractal dimension of classical \cite{S19, S21} and quantum \cite{QuantumCirculation21} circulation in turbulent flows.

These authors used a square loop $C$ with a variable side $a$ and fitted their data to constant values of $\lambda(p)$ for each $p$ over the inertial range of $\log a$, neglecting possible systematic deviations from that fit. 

In the future, it would be interesting to fit that data by our formula with the running coupling constant $g_R(\log a)$.

Here are the DNS data from these papers, fitted by our formula with constant $g_R$ Fig.\ref{fig::LambdaClassical}, Fig.\ref{fig::LambdaQuantum}.
The more detailed analysis of underlying data may reveal this slow dependence of $\log a$.

One should give proper credit to the numerical work \cite{S19, S21, QuantumCirculation21} and successful phenomenological models \cite{SY21, SY22} explaining these data by approximating the chain of the Hopf equation for velocity moments.  

However, phenomenological theory cannot distinguish between the constant fractal dimensions and those running with logarithms of scale; this is a task for a microscopic theory.

The microscopic theory so far says nothing about the transient region $p \sim 5$ where both asymptotic laws break. Only the phenomenological theory \cite{SY21, SY22} accurately describes the fractal dimensions (for velocity differences only !) in the whole domain of small and large $p$.

\section{Questions and answers}

I had to answer various questions and critical comments when discussing this paper with colleagues. 

Let us go through the most interesting questions; it would help a like-minded reader to understand this unconventional work.
\begin{enumerate}
\item Q: 

By differentiating your relation for the velocity at the center of the Burgers vortex, you get an identity
\begin{eqnarray}
   &&\pbyp{v_t(l)}{l} =\pbyp{\left(\vec v(\vec C(l))\cdot \vec C'(l)\right)}{l} = \nonumber\\
   &&\vec C'(l)\cdot \hat S(\vec C(l))\cdot \vec C'(l) +  \vec v(\vec C(l))\cdot \vec C''(l) =  \nonumber\\
   &&c(l) + \vec v(\vec C(l))\cdot \vec C''(l) 
\end{eqnarray}
From that, we conclude that 
\begin{eqnarray}
   v_t(l) = \int d l c(l) + \int d l  v_n(l) \kappa(l)
\end{eqnarray}
where $\kappa(l)$ is the local curvature of the loop, and $v_n(l)$ is the velocity projection on the inner normal of the loop.

According to the Burgers solution, the second term vanishes at the center of the loop.
Therefore, with the positive $c(l)$ you need for stability, this formula contradicts the periodicity of velocity at the loop.

A:
 I am afraid I disagree with your formula for the tangent component of the velocity field. My formula, \eqref{VelocityStrainIntegral}, is manifestly periodic (and also parametric invariant, as it should be). With explicit initial data, it reads
\begin{equation}
    \vec v(\vec C(l)) =\vec v(\vec C(0)) + \int_0^l d l' \hat S(C(l')) \cdot \vec C'(l')
\end{equation}

Using the eigenvalue condition  $\hat S(C(l')) \cdot \vec C'(l') = c(l') \vec C'(l') $ it can be written as
\begin{equation}\label{cottectEq}
    \vec v(\vec C(l)) = \vec v(\vec C(0)) + \int_0^l d l' c(l') \vec C'(l')
\end{equation}

Multiplying this equation by $\vec C'(l)$, we find the correct relation for the tangent velocity
\begin{eqnarray}
    v_t(l) = \vec v(\vec C(0)) \cdot \vec C'(l) + \int_0^l d l' c(l') \vec C'(l') \cdot \vec C'(l)
\end{eqnarray}
Take $l = L$ in \eqref{cottectEq}; we have the periodicity condition
\begin{equation}
    \vec 0 = \int_0^L d l' c(l') \vec C'(l')
\end{equation}
This expression does not change if you shift $c(l')$ by any constant, as the extra term will integrate into zero
\begin{equation}
     \int_0^L d l'  \vec C'(l') = \vec C(L) - \vec C(0) = \vec 0;
\end{equation}

In particular, one may subtract from $c(l')$ its (positive) mean value on the loop. Thus positivity of $c(l)$ does not contradict the periodicity. 

\item Q:
What is the boundary value of the Euler velocity at the boundary $C$ of the vortex surface? The normal to the loop component of velocity must vanish for mass conservation (incompressibility), which condition directs the boundary velocity along the loop. Do you agree?

A: There is no boundary value of the Euler velocity, as it diverges as $1/r$, where $r\ra 0$ is the radius of the vortex tube.

Velocity at the center of the Burgers vortex is directed along the loop, the same as vorticity, thus providing nontrivial helicity density.

This velocity component normal to the loop $v_n$ is not zero for the following reason. This normal inner component of velocity is singular (diverges as $1/r$ near the edge), so one cannot neglect the existence of a thin Burgers tube (see Fig.\ref{fig::TubeSpiral} ).

The boundary conditions for the Euler flow are imposed at the surface of the tube, not at its center, and there is a match.

\item Q:

What prevents the  Kelvinon from flying in the normal direction to the ring, as all vortex rings do?
I also wanted to ask if the Kelvinon has a non-zero impulse $\propto \int_{all space} \vec x \times \vec \omega d^3 x$.  The impulse is proportional to the net linear momentum. For the Burgers loop part, it is not difficult to compute the impulse as a line integral around the loop, which is non-zero.

A:
The Neumann boundary condition for the potential part in the velocity (with the rotational part being the \BS{} integral), by definition of the stationary state, cancels the sheet's motion (normal velocity vanishes, and tangent one does not mean a moving surface).
As for the Burgers tube attached to the surface at the edge, the same thing happens. The mean velocity in the normal plane to the ring vanishes at every point of the loop, so it does not move.

The finite negative radial strain pushes flow inside the ring, moving along the axis in the rotating vortex with circulation $\Gamma_C$. In addition, there is a (much faster) circular motion in the normal plane, going around the small circle at each point of the loop. Together, these motions create a spiral, shown in Fig. \ref{fig::TubeSpiral}.

That self-induced velocity $v =\Gamma \log R/w $ comes from the region far from this surface, just like my anomalous part in the Hamiltonian. This part, indeed, cancels by the surface contribution by the Neumann boundary conditions, which are designed for a steady state.

So, the vortex surface attached to the vortex ring acts like a parachute, or, better to say, as a wing, stopping the normal motion. The flux goes around the surface and the ring in circles, as shown in Fig.\ref{fig::EulerOutside}.

As for the net momentum, I cannot prove it is zero, even for a stationary velocity field tangent to the surface and the ring.

All I need is a stationary distribution of circulation based on stationary flow. The loop and the surface must also be stationary, but the net momentum of the flow could be finite. The integral of the velocity field over volume is not restricted by its value at the surface. Take an example of the flow around the rigid body.

I analyze this question in the section \ref{boundaryConditions}.

In that section, I also discuss the relation of the CVS conditions for the vortex sheet $\hat S \cdot \Delta \vec v =0 $ and the new boundary condition $\hat S \cdot \vec n = c \vec n$. I argue that they are related, given that the velocity gap linearly vanishes at the edge of the surface.

\item Q:

Is the diffusion of the sheet also everywhere balanced on its surface by strain presumably induced by other portions of the sheet? 

A:
This balance leads to the CVS condition, which I analyzed in my previous work. In a stationary solution, the diffusion term exactly cancels the advection term, which happens with both Burgers and Townsend solutions in the vorticity region. For the flat surface, the normal velocity vanishes, and the negative normal strain makes the solution stable.

For a curved vortex sheet, this is provided by the CVS equations. So, it does not happen for arbitrary sheets. The sheet's shape must adjust for this cancellation, which is what happens in my exact CVS solution with the hyperbolic sheet $ x |y|^\mu = const$.

As for decay, the decay time (and reconnection time) are not infinite, but they grow with the Reynolds number. Thus my "stationary" solution only holds for this growing decay time.

\item Q: 

When you compute the energy, for example, the equation with Euler's constant, do you use an expression for the energy in terms of the vorticity? I think that the expression to use is $\propto \int_{all space} \vec{B} \cdot \vec{\omega} d^3 x$, where $\vec{B}$ is the vector potential.  Is this the expression you used?  

A:
 No, why? I used ordinary $\int_V d^3 \vec r \vec v^2/2$ and substituted the Burgers solution. BTW note that the Burgers vortex may generally have a constant velocity in the $z$ direction.

Look at the computation in the appendix in my Phys Reports \cite{M23PR}. 

\item Q:

Did you prove the positivity of $c(l)$?

A:

I could not prove it in the general case, but I have presented arguments in favor in Section \ref{boundaryConditions}. It is left as an open problem to prove it.

\item Q:

I wanted to mention the Kelvin-Benjamin \textbf{maximum} energy principle: A steady inviscid vortex has maximum energy with respect to isovortical perturbations (i.e., imagine perturbing the vortex by an arbitrary incompressible velocity field acting for a short pseudo-time). Given this, shouldn't the relaxation procedure maximize the Hamiltonian?
Perhaps, this principle does not apply because the Hamiltonian is infinite in an inviscid limit?

A:

Mathematically, I am investigating the turbulent (not inviscid) limit where the Reynolds number is arbitrarily large but not infinite.

The Hamiltonian minimization is a tricky part, indeed. Consider the Euler flow outside the Burger's tube and the attached sheet. The parameters of this flow (including the normalization of the Clebsch field $Z$) minimize the Euler Hamiltonian ( the integral of the space outside the tube and the sheet), with the boundary conditions at the surface of the tune and the sheet. As I have shown, these boundary conditions involve the outside flow parameters (the boundary value of $S_3$,  the shape of the vortex sheet, and the $Z$ factor).

Now, the Euler part of the Hamiltonian (integral over the outside region) is not conserved. Its time derivative equals minus anomalous dissipation.
However, the total Hamiltonian, which includes the integral inside the vortex tube and sheet) is conserved. Its time derivative vanishes with a proper value of the normalization factor $Z$, an Euler integral of motion.

For this particular value of $Z$, which provides the energy balance, the total Hamiltonian is conserved in the NS (dissipation is exactly canceled by energy pumping).

Our Matching Principle at work here: all the variables and parameters of the Euler flow are to be defined as those of the NS flow with an infinitesimal viscosity. Logarithmic divergences prevent from setting viscosity to zero, but I generalize the Euler flow (call it turbulent flow) as the one minimizing the conserved Hamiltonian.

So, this is a definition of the inviscid limit of the NS in the presence of anomalies; this definition formally becomes the Euler theory with anomalous terms in the Hamiltonian. There is no inviscid limit, but there is an intermediate logarithmic regime where all the negative powers of the Reynolds number are neglected, but the logarithmic terms remain. These terms are summed up by the renormalization group equation, leading to asymptotic freedom.

The absolute maximum of the Hamiltonian would not be mechanically stable, plus it only exists if you impose restrictions.
In our case, there is, indeed, a restriction of a cancellation between the dissipation and pumping. 
Such a restriction would eliminate the zero and the infinite solution for parameter $Z$ in the Euler velocity field.

In that sense, I agree with your comment-- given the energy balance, this is just an extremum.

Our work belongs to theoretical physics, not applied math. Rather than solving established equations, we postulate new ones to be verified in experiments and DNS. Our new matching principle removes ambiguities from the weak Euler solutions, leading to asymptotic freedom.

Remains to be seen whether it applies to real Turbulence.
\end{enumerate}
\section{Conclusions}

Here is what we added to the previous results published in the review paper \cite{M23PR}.
\begin{itemize}
    \item We clarified the topology of the Kelvinon. Its boundary value $\phi_2(\partial \mathcal T)$ at the surface of the infinitesimal tube $\mathcal T$ surrounding the singular line $C$ maps a torus on a circle, which mapping is described by two integer winding numbers related to velocity circulations around two cycles of the torus.
    \item The 3D field $S_a(\vec r)$ maps the compactified 3-space without the infinitesimal tube $\mathcal T$ onto one of the two caps on a sphere $\bS^2$ separated by the circle $\gamma: S_3 = \const{}$.
    \item We visualized the velocity field inside and outside Kelvinon by tracking liquid particles marked by different colors. The inside flow Fig.\ref{fig::TubeSpiral} goes in spirals parallel to the loop, whereas the outside flow Fig.\ref{fig::EulerOutside} also contains the tornadoes attached to each side of the vortex sheet.
    \item We modified the energy balance analysis of \cite{M23PR} using conventional random forces and expanding the energy pumping into the Kelvinon in series in the running coupling constant $g_R \sim 1/\log R$. This approach gives us a microscopic definition and corrects the $\log R$ dependence of the phenomenological parameters in the circulation PDF tails \cite{M23PR}.
    \item We studied the vorticity field in the topological family of Kelvinons and presented a smooth \CL{} field in the Euler region (outside the vortex sheet and the tube) for arbitrary smooth loop $C$ and arbitrary winding numbers $n, m$.
    \item We found the self-consistency conditions for the Kelvinon field \eqref{SelfConsistencyGammaC}, \eqref{EigenvalueCondition} from the matching conditions with the Burgers vortex, overlooked in the \cite{M23PR}.
    \item Using these conditions, we removed the ambiguity in relative signs of the winding numbers $n, m$: they must have opposite signs.
    \item We computed and compared predictions of the Kelvinon theory for the fractal dimension $\lambda(p, \log r)$ in the leading perturbation expansion in $1/\log r$ with the DNS data (Fig.\ref{fig::LambdaClassical}, \ref{fig::LambdaQuantum}) with a good fit except for the transient region $4 < p < 7$.
\end{itemize}

In conclusion, we suggested a microscopic quantitative approach to the turbulence problem, assuming a low density of vortex structures. We presented some predictions for the multifractal indexes, modified by powers of the logarithm of the scale.

Here is what still needs to be elaborated and clarified.
\begin{itemize}
    \item The self-consistency conditions \eqref{SelfConsistencyGammaC}, \eqref{EigenvalueCondition} need to be investigated further. Presumably, the boundary values of the \CL{} field on each side of the discontinuity surface provide the set of free parameters needed to satisfy these self-consistency conditions.
    \item The notion of the region occupied by Kelvinon needs to be clarified and defined unambiguously. With correct definition, observable results should not depend upon the shape of the boundary of this region, and its volume should be a well-defined functional of the loop $C$.
    \item Higher correction in perturbation expansion in the running coupling constant $g_R$ need to be computed; fractal dimensions should become universal functions of the logarithm of scale without any phenomenological parameters to fit the DNS data.
\end{itemize}
\newpage
\section*{Acknowledgments}

I benefited from discussions of this theory with my friends and colleagues: Konstantin Khanin, Camillo De Lellis, Nader Masmoudi, Sasha Polyakov, Karim Shariff, Katepalli Sreenivasan, Grigory Volovik, Pavel Wiegmann and Victor Yakhot. 

I am also grateful to Kartik Iyer and Juan Ignacio Polanco for providing me with some data from their DNS of classical and quantum Turbulence.

This research was supported by a Simons Foundation award ID $686282$ at NYU Abu Dhabi.

\section*{Data Availability}
Data sharing does not apply to this article, as no new data were created or analyzed in this study.
\newpage
\section*{References}
\bibliography{bibliography}
\newpage
\appendix
\setcounter{section}{-1}
\section{Topological family of the Kelvinon fields}\label{ExampleClebsch}

Let us present an explicit example of the \CL{} field with the required topology, which could serve as initial data for the Hamiltonian minimization by relaxation.

We introduce a surface of the minimal area bounded by our loop $C$
\begin{eqnarray}
   S_{min}(C) = \argmin_{S: \partial S = C} \int_S  d S 
\end{eqnarray}
For every point $\vec r \in R_3$, there is the nearest point $\vec r_1$ at the minimal surface $S_{min}(C)$. 
\begin{eqnarray}
   \vec r_1 = \argmin_{ \vec r' \in S_{min}(C)} (\vec r - \vec r')^2;
\end{eqnarray}

For this point $\vec r_1 $ at the surface there is also a nearest point $\vec r_0$ at its edge $C$, minimizing the geodesic distance  $d( a,  b)$ along the surface from $\vec r_1$ to the edge
\begin{eqnarray}
   &&s_0 = \argmin_{s} d(\vec r_1, \vec C(s));\\
   && \vec r_0 = \vec C(s_0);
\end{eqnarray}

Let us also introduce the local frame with vectors $\vec t(s), \vec n(s), \vec \sigma(s)$ at the loop:
\begin{subequations}
\begin{eqnarray}
&& \vec C'(s)^2 =1 ;\\
&&\vec t(s) = \vec C'(s);\\
&& \vec n(s) = \frac{ \vec C''(s)}{|\vec C''(s)|}\\
&&\vec \sigma(s) =\vec t(s) \times \vec n(s);
\end{eqnarray}
\end{subequations}\label{frame}

Our field is then defined as follows:
\begin{subequations}\label{alphabetarho}
\begin{eqnarray}
&& \alpha = \frac{2 \pi s_0}{ \oint | d \vec C|};\\
&& \beta = \arg \left((\vec r - \vec r_0)\cdot \left(\vec n(s_0) + \imath\vec \sigma(s_0)\right) \right);\\
&& \rho = \sqrt{(\vec r - \vec r_1)^2 + d(\vec r_1, \vec r_0)^2} ;\\
\label{thetaEq}
&& \theta =f(\rho^2);\\
&& f(0) = \lambda;\;f(\infty) = \pi;\; f'(\rho^2) >0;\\
&& \phi =  m \alpha +  n \beta;\\
&&\vec S =  \left(\sin\theta \cos\phi, \sin\theta \sin\phi, \cos\theta\right) ;
\end{eqnarray}
\end{subequations}

In the variational solution for the Kelvinon flow, one may use this Anzatz and optimize the smooth monotonous function $f(\rho^2)$ to reach the minimum of the Euler Hamiltonian at fixed winding numbers $n, m$.

In this example, the \CL{} field maps the physical space $\mathcal G = \bR^3 \setminus\mathcal T$ onto the disk on a 2-sphere $\Omega_+ : S_3 < \cos \lambda$.
With the function $f(\rho^2)$  monotonously \textit{decreasing} from $f(0) =\lambda $ to $f(\infty) = 0$, this Kelvinon would map the physical space to the complementary region $\Omega_-$.

When the point $\vec r $ approaches the nearest point $\vec r_1$ at the minimal surface, the difference $ \vec \eta = \vec r - \vec r_1$ is normal to this surface.

When the point $\vec r_1$ approaches the nearest point $\vec r_0$ at the edge $C$ of the surface, the geodesic becomes a straight line in $R_3$, tangent to the surface and $\rho$ becomes Euclidean distance to the loop
\begin{eqnarray}
  &&d(\vec r_1, \vec r_0) \ra |\vec r_1 - \vec r_0|;\\
   &&\rho^2 \ra  |\vec r - \vec r_1|^2 + |\vec r_1 - \vec r_0|^2 = |\vec r - \vec r_0|^2 
\end{eqnarray}

Note that all variables $s_0, \vec r_0, \vec r_1, \alpha, \beta,\rho,\theta, \phi$ depend on $\vec r \in R_3$ through the minimization of the distance to the surface and the loop. By construction, $\rho= |\vec r - \vec r_0|$ away from the surface, when $\vec r_1 = \vec r_0, d(\vec r_1, \vec r_0) =0$.

The Euler angles $\theta, \phi$ for the \CL{} field take the boundary values at the loop :
\begin{eqnarray}
&& \phi(\vec r \ra C) \ra  m \alpha + n \beta;\\
&&\theta(\vec r \ra C) =  \lambda + \MO{(\vec r-C)^2};
\end{eqnarray}
and $\theta(\infty) = 0 \textit{  or }\pi$. 

Assuming the decay $f(\rho^2) \ra f(\infty) + \const{}/\rho^2$, one may estimate the decay rate of $\vec \nabla \cos\theta \sim 1/|\vec r|^3, \vec \nabla \phi\sim 1/|\vec r|$, which corresponds to vorticity decaying as $1/|\vec r|^4$.
This decay is sufficient for the convergence of the enstrophy integral at infinity.

The \BS{} integral for velocity corresponding to such vorticity would decay as $1/\vec r^2$ or faster, sufficient for convergence of the Euler Hamiltonian at infinity.
   
Let us move $\vec r$ along the normal from the surface at $\vec r_1$. Our parametrization of $\theta$ does not change in the first order in normal shift $\vec \eta = \vec r - \vec r_1$, as $\rho^2$ has only quadratic terms in $\vec \eta$.

 We conclude that the normal derivative of the \CL{} field $\phi_1 = Z(1+ \cos\theta)$ vanishes
\begin{eqnarray}
\partial_n \phi_1=0;
\end{eqnarray}

 We requested vanishing normal velocity at the discontinuity surface for this surface to be stationary.

 In terms of the \CL{} parametrization, the normal velocity would vanish provided
 \begin{eqnarray}\label{dnS3}
\partial_n \Phi = \left(\vec \nabla \times \Psi\right)_n; \; \vec r \in S_C\setminus C
\end{eqnarray}

As for the angular field $\phi$ in \eqref{alphabetarho}, its normal derivative does not vanish in the general case. The angle $\alpha$ does not change when the point $\vec r$ moves in normal direction  $\vec N(\vec r_1)$ from the surface projection $\vec r_1$ by infinitesimal shift $\vec \eta = \eps \vec N(\vec r_1) $. However, another angle $\beta$ changes in the linear order in $\eps$ as
\begin{eqnarray}
   \vec N(\vec r_1) \cdot \left(\vec n(s_0) + \imath\vec \sigma(s_0)\right) \neq 0
\end{eqnarray}

\end{document}